\documentclass[usenatbib]{mn2e}
\newcommand{\bib}{\bibitem[\protect\citeauthoryear}
\usepackage{epsf}
\begin{document}
\title[Jet deceleration]{Dynamical models for jet deceleration 
in the radio galaxy 3C\,31}
\author[R.A. Laing \& A.H. Bridle]{R.A. Laing
       \thanks{E-mail: rlaing@astro.ox.ac.uk}$^{1,2}$, A.H. Bridle$^3$\\
     $^1$ Space Science and Technology Department, CLRC, 
     Rutherford Appleton Laboratory,
     Chilton, Didcot, Oxfordshire OX11 0QX \\
     $^2$ University of Oxford, Department of Astrophysics, Denys
     Wilkinson Building, Keble Road, Oxford OX1 3RH \\
     $^3$ National Radio Astronomy Observatory, 520 Edgemont Road, 
     Charlottesville, VA 22903-2475, U.S.A.}

\date{Received }
\maketitle

\begin{abstract} 
We present a dynamical analysis of the flow in the jets of the low-luminosity
radio galaxy 3C\,31 based on our earlier geometrical and kinematic model
\citep{LB02} and on estimates of the external pressure and density distributions
from {\em Chandra} observations \citep{Hard_3C31}. We apply conservation of
particles, energy and momentum to derive the variations of pressure and density
along the jets and show that there are self-consistent solutions for
deceleration by injection of thermal matter.  We initially take the jets to be
in pressure equilibrium with the external medium at large distances from the
nucleus and the momentum flux to be $\Pi = \Phi / c$, where $\Phi$ is the energy
flux; we then progressively relax these constraints. With our initial
assumptions, the energy flux is well determined: $\Phi \approx$ 9 -- 14 $\times
10^{36}$\,W.  We infer that the jets are over-pressured compared with the
external medium at the flaring point (1.1\,kpc from the nucleus) where they
start to expand rapidly.  Local minima in the density and pressure and maxima in
the mass injection rate and Mach number occur at $\approx$ 3\,kpc.  Further out,
the jets decelerate smoothly with a Mach number $\approx$ 1.  The mass injection
rate we infer is comparable with that expected from stellar mass loss throughout
the cross-section of the jet close to the flaring point, but significantly
exceeds it at large distances.  We conclude that entrainment from the galactic
atmosphere across the turbulent boundary layer of the jet is the dominant mass
input process far from the nucleus, but that stellar mass loss may also
contribute near the flaring point.  The occurrence of a significant
over-pressure at the flaring point leads us to suggest that it is the site of a
stationary shock system, perhaps caused by reconfinement of an initially free
jet.  Our results are compatible with a jet consisting of $e^-e^+$ plasma on
parsec scales which picks up thermal matter from stellar mass loss to reach the
inferred density and mass flux at the flaring point, but we cannot rule out an
$e^-p^+$ composition with a low-energy cut-off.
\end{abstract}


\section{Introduction}
\label{Introduction} 

The measurement of basic flow variables such as velocity, pressure and
density in extragalactic radio jets has proved to be an intractable
problem, most of the estimates in the literature being highly
model-dependent (see \citealt{Leahy91} for a review).   We have recently
shown that the total and polarized emission of the inner jets in the
nearby radio galaxy 3C\,31 can be modelled accurately on the assumption
that they are symmetrical, axisymmetric, relativistic, decelerating flows
\citep{LB02}, and we derived a kinematic model for the jet flow that
combined longitudinal deceleration and a transverse velocity gradient.

In order to make further progress in understanding jet dynamics, we need a
physical model for the deceleration process. Mass loading must occur, but
without disruption of the flow. As \citet{Beg82} first pointed out, a jet
can decelerate without being completely decollimated, but only in the
presence of an external galactic pressure gradient, which effectively
transforms heat back into kinetic energy.  It is not straightforward to
estimate the mass input from observations: synchrotron emission gives no
direct evidence for the jet composition on kpc scales, and constraints
from Faraday rotation are weak. Two principal mechanisms have
been proposed for mass loading:
\begin{enumerate}
\item
injection from stellar winds within the volume that is traversed
by the jet \citep*{Phi83,Kom94,BLK96}, 
and
\item
entrainment from the galactic atmosphere across an unstable boundary
layer, and subsequent communication with the rest of the jet through
ingestion of the thermal material and viscous interactions
\citep{Baa80,Beg82,Bic84,Bic86,DeY96,RHCJ99,RH00}.
\end{enumerate}
In the remainder of this paper, we will refer to these processes as {\em
internal} and {\em external} entrainment, respectively.

The majority of theoretical work in the literature concerns
non-relativistic jets, but there have been two approaches to the
quantitative study of relativistic jet deceleration: through analytical
models and simulations \citep{Kom94,BLK96} and through conservation law
analysis \citep{Bic94}. \citet{Kom94} considered analytically the case of
an electron-positron jet decelerating as a result of internal entrainment 
and \citet{BLK96} made numerical simulations of decelerating
electron-proton jets. Both of these references assumed that the jet
dynamics were dominated by thermal particles (with energies too low to be
seen via synchrotron radiation), although some of the cases they
considered were hot enough to have a relativistic equation of state.
These calculations were not designed to be compared directly with
observations of individual objects and are restricted to internal
entrainment.

\citet{Bic94} used the laws of conservation of mass, momentum and energy
in a quasi-one-dimensional approximation to demonstrate the feasibility of
deceleration from pc to kpc scales for relativistic jets, considering two
specific sources: NGC\,315 and NGC\,6251. In contrast to \citet{Kom94} and
\citet{BLK96}, Bicknell assumed that relativistic particles are
energetically dominant, and therefore that an ultra-relativistic equation of
state is appropriate throughout. His formulation is general enough to
cover both internal and external entrainment.

Our model for the jets in 3C\,31 provides one essential ingredient for a
dynamical analysis -- the velocity field -- but we also need to estimate
how much mass participates in the flow.  Our solution for the jet
kinematics can be used to constrain the mass flux using the
conservation-law formalism of \citet{Bic94}, but only if we also have an
accurate prescription for the external pressure and density.  Such a
prescription has recently been derived from {\em Chandra} observations by
\citet{Hard_3C31}, and the present paper describes the resulting dynamical
analysis of jet deceleration in 3C\,31.

The conservation-law approach is described in Section~\ref{Model}.  The 
results are presented in Section~\ref{results} and are discussed in the
context of theoretical models in Section~\ref{discussion}. Our conclusions
are summarized in Section~\ref{conclusions}.

Throughout this paper, we adopt a Hubble constant $H_0$ =
70\,km\,s$^{-1}$\,Mpc$^{-1}$.  We take the redshift of NGC\,383 (the
parent galaxy of 3C\,31) to be 0.0169; this is the mean of values from
\citet{Smith2000}, \citet{HVG} and \citet{RC3}.  The resulting conversion
factor between angular and linear size is 0.34\,kpc /arcsec.  We refer to
two quantities that are conventionally notated as $\beta$.  We use $\beta$
alone for the normalized velocity $v/c$, and $\beta_{\rm atm}$ for the
form parameter in models of hot galactic atmospheres.

\section{Conservation-law analysis}
\label{Model}

\subsection{Overview}

Our analysis uses conservation of particles, energy and momentum, and is
based on that of \citet{Bic94}, but with some important differences:
\begin{enumerate}
\item We take the angle to the line of sight, velocity, and area of the
jets from our kinematic model, thereby removing a number of free parameters.
\item We do not need to make any assumptions about the
relation between the internal pressure $p$ of the jet and the synchrotron
minimum pressure $p_{\rm sync}$, other than that $p \geq p_{\rm sync}$.
\item We infer the density and pressure distributions of the surrounding
medium from X-ray observations.
\item The effects of
buoyancy are significant for our analysis, and we include them.
\end{enumerate}
The analysis remains quasi-one-dimensional, in the sense that we consider only
the component of velocity along the jet axis, but we allow the jet's
cross-sectional area to vary.  We justify this assumption by noting that the
maximum angle between the flow direction and the jet axis is 19$^\circ$. The
axial velocity component is $>$0.945 of the total velocity even at the edge of
the jet in the region where the lateral expansion is most rapid.  The error in
the quasi-one-dimensional approximation is therefore small compared with the
uncertainties we consider below.

\subsection{Geometry: the three jet regions}

\begin{figure}
\epsfxsize=8.5cm
\epsffile{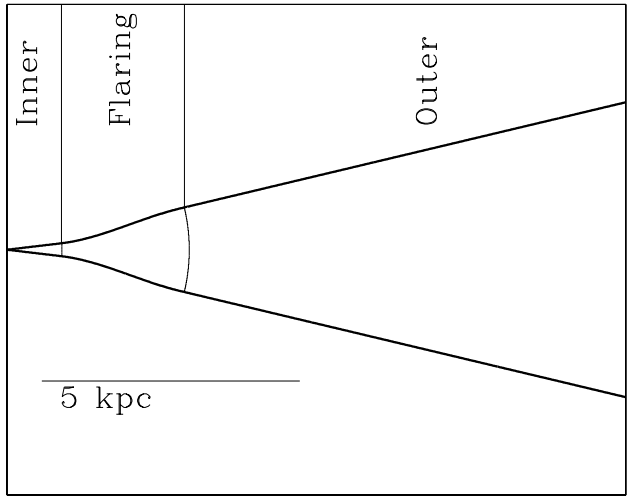}
\caption{The model geometry for the jets in 3C\,31, showing the three
regions.\label{geom}} 
\end{figure}

\begin{figure}
\epsfxsize=8.5cm
\epsffile{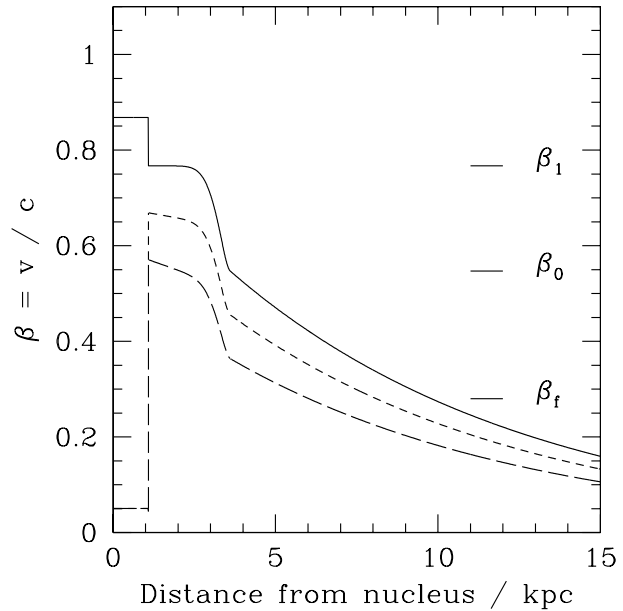}
\caption{Profiles of the velocity along streamlines. Full line: on-axis; short
dash: intermediate; long dash: jet edge.  The on-axis profile in our region of
interest is parameterized by the values at the flaring point ($\beta_1 =
0.767$), the boundary between the flaring and outer regions ($\beta_0 = 0.547$)
and a point at 9.6\,kpc from the nucleus in the outer region ($\beta_f =
0.280$).  In addition, the velocity exponent $H$ (see \citealt{LB02} for more
details) governs the steepness of the deceleration at the end of the flaring region.
\label{vel-profile}}
\end{figure}

The essential features needed from the kinematic model \citep{LB02} are the
geometry and velocity field of the jets.  We divide the jets of 3C\,31
into three regions by the shapes of their outer isophotes
(Fig.~\ref{geom}; see \citealt{LB02} for more details). The regions also
turn out to be distinct in their kinematic structures
(Fig.~\ref{vel-profile}).  We have used the fitted angle to the line of
sight (52.4$^\circ$) to correct for projection, and all distances are
given in a plane containing the jet axis. The regions are:
\begin{description}
\item [{\bf Inner}] (0 to 1.1\,kpc): a cone of intrinsic
half-angle 6.7$^\circ$ in which the fitted central velocity is 0.8 --
0.9$c$. There is no unambiguous evidence for deceleration, but a
significant component of emission appears to come from slow-moving
material.   
\item [{\bf Flaring}] (1.1 to 3.5\,kpc): the outer isophotes spread
rapidly and then recollimate. The boundary with the inner region appears
to mark a discontinuity at which the emissivity increases abruptly and the
velocity probably decreases. In this region, the jets decelerate rapidly
after an initial slow decline in velocity.
\item [{\bf Outer}] (3.5 to 12\,kpc): the jets continue to expand
in a cone of intrinsic half-angle 13.1$^\circ$ and decelerate smoothly.
\end{description}
We refer to the boundary between the inner and flaring regions as the
``flaring point'', following \citet{Parma87}.  We do not attempt to apply
the conservation-law approach to the inner region, for several reasons.
Firstly, the jets are weak and poorly resolved close to the nucleus, so
our conclusions about their velocity structure are tentative. Secondly,
our formalism cannot handle the inferred discontinuity in velocity and
pressure at the flaring point without introducing extra
complexity. Finally, the external pressure and density are uncertain at
very small distances from the nucleus, where the X-ray emission from core,
jet and hot gas is unresolved by {\em Chandra}. In particular, the
presence of an unresolved, dense component of hot gas was postulated by
\citet{Hard_3C31} in order to explain the observed X-ray spectrum of the 
core. We restrict our conservation-law analysis to between 1.1 and 12\,kpc
from the nucleus.

In what follows, $r$ is a radial distance from the nucleus and $A(r)$ is
the cross-sectional area of the jets derived from our geometrical model.

\subsection{Velocity profile}

In order to reduce the problem to a quasi-one-dimensional form we need to
average in some way over the transverse jet velocity profile. 
For the flaring and outer regions, the best-fitting transverse profile 
remains remarkably constant: the velocity at the edge of the jet is
$\approx$0.7 of its central value throughout these regions. The transverse
variation of emissivity is also modelled, and may be used to derive the
corresponding relativistic particle and energy density profiles, but only
if some assumption is made about the relative energies in field and
particles.  We do not know  the spatial distribution of the entrained
material and would expect qualitative differences between
stellar mass loss (which would cause material to be injected fairly
uniformly) and entrainment around the edges of the jets. 

We assume that all of these complications can be absorbed into a single
velocity function $\beta(r)c$, which is bounded at any value of $r$ by the
maximum (centre) and minimum (edge) values of our kinematic model.  The
approximation is reasonable, since we infer a velocity difference of only
$\approx$30\% between centre and edge of the jet and, in order to slow the
jet effectively, any entrained material must become well mixed with the
relativistic particles and field. We have experimented with a number of
expressions for $\beta(r)$ consistent with this constraint.

The widest range of acceptable solutions (in the sense defined below) is given
by a simple law in which the velocity is a constant fraction $f$ of the on-axis
value for the kinematic model (the full line in Fig.~\ref{vel-profile});
hereafter a ``constant-fraction'' profile.  We have also investigated ``ramped
profiles'' in which the velocity decreases from the on-axis value, $\beta_c(r)$,
at the flaring point ($r = r_1$) to a fraction $f$ of the on-axis value at the
end of the modelled region ($r = r_{\rm max}$).  We present results for a linear
variation: $\beta(r) = \beta_c(r) [1 + (f-1)(r - r_1)/(r_{\rm max} - r_1)]$.
Other functional forms give essentially identical results for a given value of
$f$.  For both constant-fraction and ramped profiles, we have investigated the
range $0.6 \leq f \leq 1$, the lower bound being slightly less than the
fractional velocity at the edge of the jet in our kinematic model.

\subsection{Conservation laws}

Following \citet{Bic94}, we make the approximation that the jet material
has an ultra-relativistic equation of state throughout the modelled region
and therefore that dissipation of bulk kinetic energy results only in an
increase in the internal energy, $u$, of relativistic particles and
magnetic field, with $p = u/3$. We also assume that the entrained enthalpy
is negligible.  We have verified post hoc that an extremely small fraction
of the internal energy of the jet is lost via synchrotron radiation in the
modelled region.  We take the quantities $\Phi$ (the energy flux, with the
contribution from rest mass subtracted) and $\Pi$ (the momentum flux) to
be conserved. From equations (18) and (27) of \citet{Bic94}, and after
making the quasi-one dimensional approximation for the buoyancy term, we
have
\begin{eqnarray}
\Phi & = & [(\Gamma^2-\Gamma)\rho c^2 +4\Gamma^2 p]\beta c A \label{E-flux} \\ 
\Pi & = & [\Gamma^2\beta^2(\rho c^2 + 4p) + p - p_{\rm ext}] A \nonumber \\ 
& + &\int_{r_1}^r  A \frac{dp_{ext}}{dr} \left [1 -
\frac{\Gamma^2(\rho c^2 + 4p)}{c^2(1 + \beta^2)\rho_{\rm ext}}
\right] dr \label{mom-flux} \\ \nonumber
\end{eqnarray}
Here, $p$ and $\rho$ are the internal pressure and density of the jet,
$p_{\rm ext}$ and $\rho_{\rm ext}$ the pressure and density of the
external medium and $\Gamma$ the Lorentz factor of the bulk flow.  $r =
r_1$ at the innermost modelled point (the flaring point). Note that we are
implicitly assuming that the kinetic energy associated with turbulence is
small compared with that of the bulk flow.  Unlike \citet{Bic94}, we
cannot neglect buoyancy (the second term in the equation for momentum
flux), as the solutions turn out to require transonic flow over large
distances.

The unknowns at each position are the internal pressure and density.  If
the external pressure and density are known, then for given values of the
energy and momentum fluxes $\Phi$ and $\Pi$ we can solve
equations~(\ref{E-flux}) and (\ref{mom-flux}) for the run of internal density
 and pressure along the jets. In the absence of buoyancy, there
is an  algebraic solution. If the buoyancy term is significant, we
can solve the equations using a simple iterative scheme at a grid of
locations from the flaring point outwards.  This is necessary because the
buoyancy integral depends on the density and pressure gradient at the
current point (equation~\ref{mom-flux}).

We can then derive 
the generalized internal Mach number
\citep{Kon80}: 
\begin{equation}
{\cal M} = \frac{\Gamma \beta}{\Gamma_s \beta_s}
\end{equation}
where
$\beta_s c$ 
is the internal sound speed,
\begin{equation}
\beta_s^2 = \frac{4p}{3(\rho c^2 + 4p)} 
\end{equation} 
\citep{Bic94} and $\Gamma_s =
(1-\beta_s^2)^{-1/2}$.  
The ratio of
rest-mass energy to enthalpy, again as defined by \citet{Bic94}, is:
\begin{equation}
{\cal R} = \frac{\rho c^2}{u + p} = \frac{\rho c^2}{4p} \label{R-eqn}
\end{equation}
and the mass flux:
\begin{equation}
\Psi = \Gamma \rho \beta c A 
\end{equation}

In order to reduce the number of unknown parameters and hence to derive a unique
solution of the conservation-law equations, we initially take $\Phi = \Pi c$.
This is likely to be a very good approximation if $\beta$ is close to 1 on pc
scales, as required by Unified Models and by the observation of superluminal
motion on parsec scales in FR\,I jets \citep{UP95,Giov}.  The effects of
dropping this assumption are investigated in Section~\ref{mom-flux-var}.  We
also test the self-consistency of our other assumptions and explore the effects
of relaxing them in Section~\ref{vars}.

\subsection{External density and pressure}

Our estimates of the surrounding density and pressure profiles are taken
from \citet{Hard_3C31}. The density profile is
the sum of two beta models, for the gas associated with the central galaxy
NGC\,383 and with the surrounding group, respectively.
\begin{eqnarray}
n_{\rm ext}(r) & = & n_c(1 + r^2/r_c^2)^{-3\beta_{\rm atm,c}/2} \nonumber\\ 
               & + & n_g(1 + r^2/r_g^2)^{-3\beta_{\rm atm,g}/2} \\ \nonumber
\end{eqnarray}
The temperature is taken to be:
\begin{eqnarray}
T & = & T_c + (T_g - T_c)\frac{r}{r_m} \mbox{~~~($r < r_m$)} \nonumber\\
T & = & T_g \mbox{~~~~~~~~~~~~~~~~~~~~~~~($r \geq r_m$)} \\ \nonumber
\end{eqnarray}
with $r_m =$ 22.8\,arcsec (7.8\,kpc).
The pressure is calculated according to the expression given by
\citet{BW93}:
\begin{equation}
p_{\rm ext} = \frac{kT(r)}{\mu X}n_{\rm ext}(r) 
\end{equation}
where $\mu = 0.6$ is the mass per particle in a.m.u. and $X = 0.74$ is the
abundance of hydrogen by mass. 

\begin{table*}
\caption{Parameterization of the external density and temperature
distributions.\label{ext-table}}
\begin{tabular}{lllll}
Component & Central density & Form factor & Core radius & Temperature \\ 
&&&&\\
\hline
&&&&\\
Galaxy & $n_c =1.8 \times 10^5$\,m$^{-3}$ & $\beta_{\rm atm,c} = 0.73$ &
$r_c =$ 1.2\,kpc  & $T_c =$ 4.9 $\times$ 10$^6$\,K\\  
&&&&\\
Group  & $n_g = 1.9 \times 10^3$\,m$^{-3}$ & $\beta_{\rm atm,g} = 0.38$ & $r_g =$
52\,kpc & $T_g =$ 1.7 $\times$ 10$^7$\,K\\ 
&&&&\\
\hline
\end{tabular}
\end{table*}

\begin{figure}
\epsfxsize=8.5cm
\epsffile{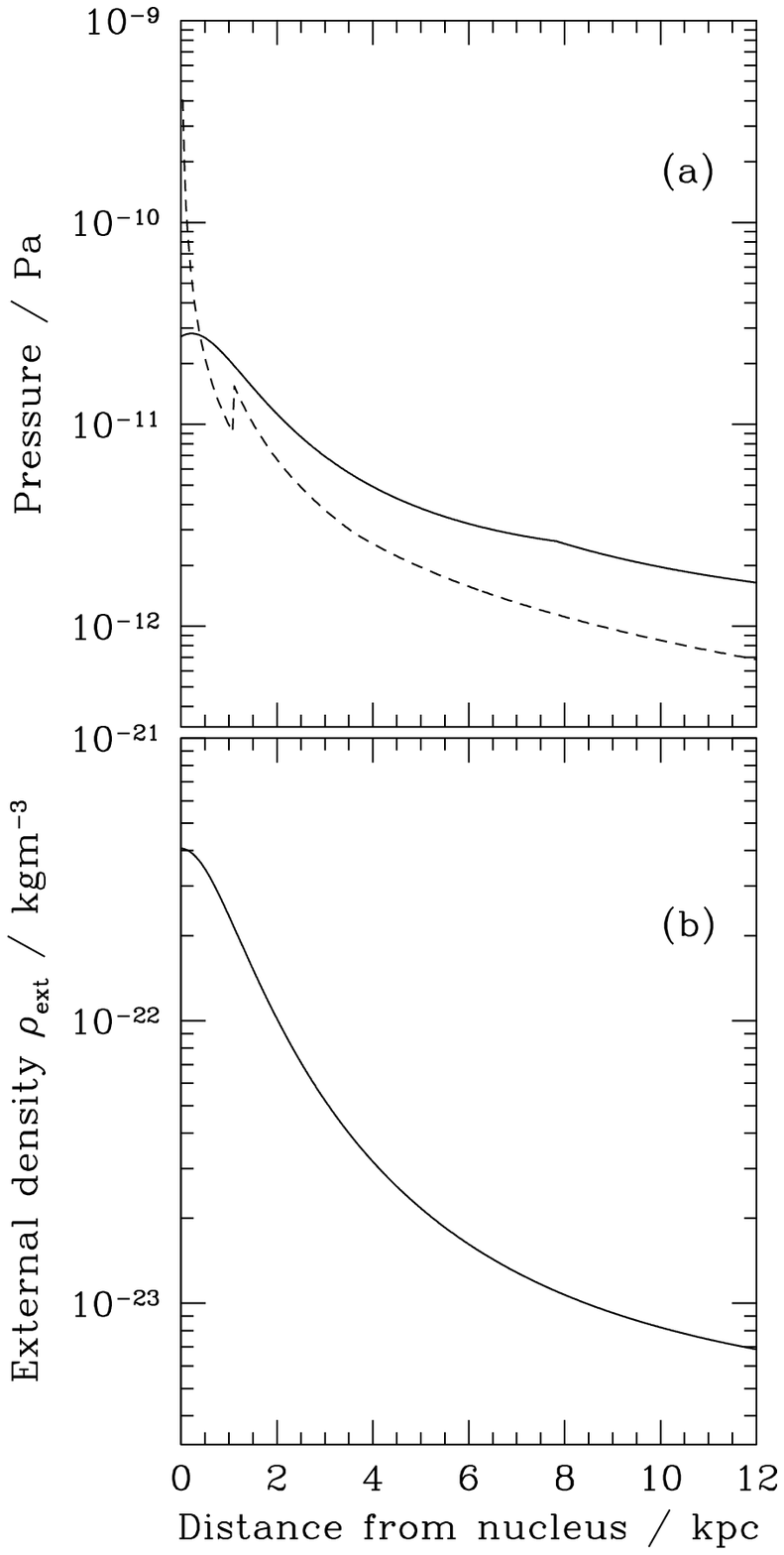}
\caption{The variations of external (a) pressure and (b) density  with
radius, as derived from {\em Chandra} observations, are shown by full
lines. The dashed curve in panel (a) represents the synchrotron minimum
pressure $p_{\rm sync}$ derived from the emissivity model of \citet{LB02}
using equation~(\ref{eq-pmin}). Although the jet appears to be
over-pressured for the first 0.5\,kpc, the presence of an
additional unresolved component of hot gas cannot be ruled out
\citep{Hard_3C31}.\label{ext-fig}}
\end{figure}

The numerical values of the parameters are listed in Table~\ref{ext-table}
and the variations of $p_{\rm ext}$ and $\rho_{\rm ext}$ with radius are shown
in Fig.~\ref{ext-fig}.  Note the 
discontinuity in the temperature gradient $dT/dr$ and hence in the
pressure gradient $dp/dr$. The latter appears in the buoyancy term of
equation~(\ref{mom-flux}) and therefore propagates into various derived
quantities, most obviously the entrainment rate (see below).

\subsection{Selection of acceptable solutions}
\label{accept}

We aim to determine what, if any, physically-reasonable solutions
exist. All self-consistent solutions must have:
\begin{enumerate}
\item $\rho > 0$  everywhere;
\item $p \geq p_{\rm sync}$, where $p_{\rm sync}$ is the synchrotron 
{\it minimum} pressure;
\item the mass flux increasing monotonically outwards;
\item a convergent iterative solution for the buoyancy integral.
\end{enumerate}
We also investigate the effects of additional constraints on the ratio of internal to
external pressure:
\begin{enumerate}
\setcounter{enumi}{4}
\item the jet pressure never exceeds that of the external medium by more  
than a factor of 10;
\item the mean ratio of the internal and external pressures $<p/p_{\rm
ext}>$ is in the range 0.5 -- 2 in the outer region.
\end{enumerate}
The over-pressure condition (v) can be justified qualitatively, since the
jets cannot be in free expansion over the whole of the modelled region
(otherwise they would expand with constant opening angle), and any
high-pressure region must be localised and not too over-pressured for the
jet to retain its collimation.  The numerical over-pressure factor is
difficult to assess, but values of $p/p_{\rm ext}$ up to 12.5 have been
suggested for reconfining non-relativistic jets \citep{Falle87} and 
extremely high pressure jumps can occur across relativistic shocks
\citep{BB}.  Condition (vi) is derived from the supposition that the outer
region must be in at least approximate pressure equilibrium.  We have
looked for solutions with and without the pressure constraints.

We have chosen to use the on-axis value of the model emissivity
$\epsilon(r)$ \citep{LB02} to calculate a representative minimum synchrotron pressure
($\epsilon$ varies across the jets, so some sort of averaging is
necessary).  Following \citet{Bic94}, we derive the pressure assuming
constant energy limits for a power-law spectrum of radiating particles
$n(\gamma) d\gamma \propto \gamma^{-(2\alpha + 1)} d\gamma$, where
$\gamma$ is the individual electron Lorentz factor and $\gamma_{\rm min}
\leq \gamma \leq \gamma_{\rm max}$.
\begin{eqnarray}
\lefteqn{p_{\rm sync}  =  \frac{1}{6\mu_0}\left(\frac{3+\alpha}{1+\alpha}\right)} \nonumber \\
             & \times &
\left [\epsilon \mu_0 \left(\frac{1+\alpha}{2\alpha-1}\right)(m_e c^2)^{1-2\alpha}
(\gamma_{\rm min}^{1-2\alpha}-\gamma_{\rm
max}^{1-2\alpha})\right]^\frac{2}{3+\alpha}\label{eq-pmin}\\ \nonumber
\end{eqnarray}
where $m_e$ is the electron mass and $\mu_0$ is the permittivity of free space.
We also assume that there are no relativistic protons and that the filling
factor is unity, to ensure that $p_{\rm sync}$ is a lower limit to the correct
value.  Initially, we take $\gamma_{\rm min} = 10$ and $\gamma_{\rm max} =
10^5$.  The upper limit would correspond to an emitted frequency of $\sim 2
\times 10^{12}$\,Hz for a minimum-pressure magnetic field at the flaring point,
roughly consistent with the synchrotron break frequency derived from the radio
-- X-ray spectrum of the inner and flaring regions by \citet{Hard_3C31}.  We
have no direct evidence for emission at $\nu \gg$ 10\,GHz in the outer region,
however.  The lower limit to the Lorentz-factor distribution is harder to
estimate. Fortunately, $\alpha = 0.55$ for the jets in 3C\,31, so the dependence
of $p_{\rm sync}$ on the energy limits is very weak, at least if the form of the
spectrum persists to low energies.  We demonstrate the effect of changing the
energy limits in Section~\ref{sync-press-var}.

Our approach is to calculate models for a range of energy flux, $\Phi$,
determining which solutions are acceptable. We start with solutions in
which the jets are close to pressure equilibrium at large distances from
the nucleus, and progressively relax our assumptions.

\section{Results}
\label{results}

\subsection{Reference model}
\label{refmod}

A priori, the most physically plausible solutions are those which are as
close as possible to pressure balance in the outer region.  Although the
shape of the jet in this region is conical, consistent with free expansion of an
over-pressured flow, such solutions are not self-consistent:
\begin{enumerate}
\item We infer Mach numbers $\cal M$ \la 2 at the end of the flaring
region, implying a half-opening angle \ga ${\cal M}^{-1} \approx$
30$^\circ$, at least a factor of two larger than is observed.
\item We will show that the entrainment rate needed to slow the jet is
too large to be provided by stellar mass loss alone, implying that
external entrainment dominates. This would be inconsistent with 
free expansion.
\end{enumerate}
In what follows, we therefore adopt as our reference model the unique
pressure-matched solution with $\Phi = \Pi c$ and a constant velocity fraction
$f = 0.85$. This is close to the median of the transverse profile estimated by
\citealt{LB02} and is therefore consistent with the average velocity that we
infer for relativistic particles and field. An intermediate (perhaps somewhat
lower) value of $f$ is also appropriate for the entrained material, which is
likely to be concentrated at the edges of the jet, or distributed approximately
uniformly, rather than being concentrated towards the axis.  This model has an
energy flux $\Phi = 1.1 \times 10^{37}$\,W and demonstrates the qualitative
properties common to all acceptable solutions.  In Section~\ref{vel-var-peq}, we
examine pressure-matched solutions with other velocity laws and
Section~\ref{vars} explores the effects of relaxing the assumptions on pressure
balance and momentum flux.

\subsubsection{Pressure}

\begin{figure}
\epsfxsize=8.5cm
\epsffile{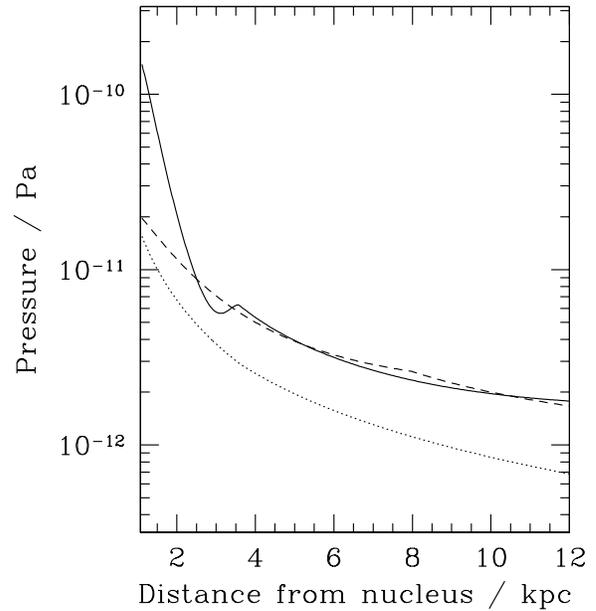}
\caption{The 
internal (full), external (dashed) and synchrotron minimum (dotted) pressures for the
reference model.\label{pressure-ref}}
\end{figure}

\begin{figure}
\epsfxsize=8.5cm
\epsffile{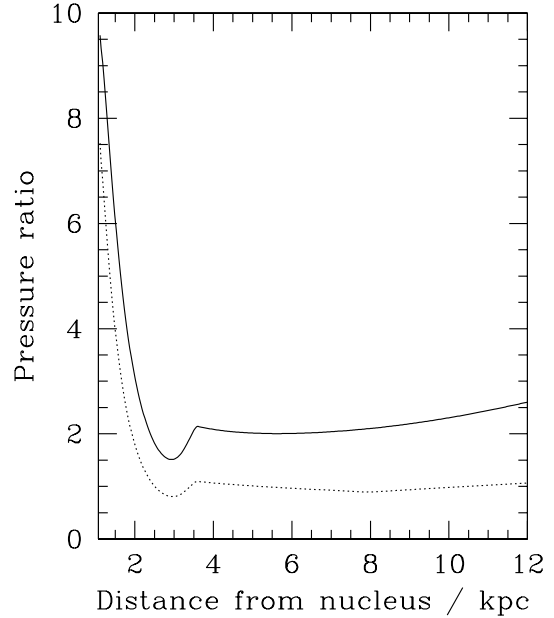}
\caption{The pressure ratios $p/p_{\rm sync}$ (full) and $p/p_{\rm ext}$
(dotted) for the reference model.\label{pratio-ref}}
\end{figure}

Fig.~\ref{pressure-ref} shows the derived internal, external and
synchrotron minimum pressures for the reference model and
Fig.~\ref{pratio-ref} shows the ratios $p/p_{\rm sync}$ and $p/p_{\rm
ext}$.  The jet is over-pressured by a factor $\approx$8 at the start of
the flaring region, comes into pressure equilibrium at 2.5\,kpc from the
nucleus, is slightly under-pressured at the end of the flaring region and
thereafter tracks the external pressure accurately.  The local pressure
minimum in the outer part of the flaring region has $p < p_{\rm ext}$,
implying that the rapid outward decrease of pressure overshoots and then
recovers to track the external value.  Although we have deliberately
chosen the solution closest to pressure equilibrium, the similarity of the
functional forms of the internal and external pressures in the outer
region is still significant.

We find that $p/p_{\rm sync} \approx 2$ except near the flaring point
(Fig.~\ref{pratio-ref}).  If the pressure is contributed primarily by
relativistic particles and magnetic field, as we suppose, then the two
must be close to energy equipartition in the outer region.  At the flaring
point, $p/p_{\rm sync} \approx 10$, so there could be a significant
departure from equipartition.  This idea could, in principle, be tested by
X-ray observations of inverse Compton emission, but the fluxes predicted
for equipartition magnetic fields are too low to be detected by {\em
Chandra} (\citealt{Hard_3C31} conclude that the X-rays observed from the
inner and flaring regions are produced by the synchrotron mechanism).

\subsubsection{Density}  

\begin{figure}
\epsfxsize=8.5cm
\epsffile{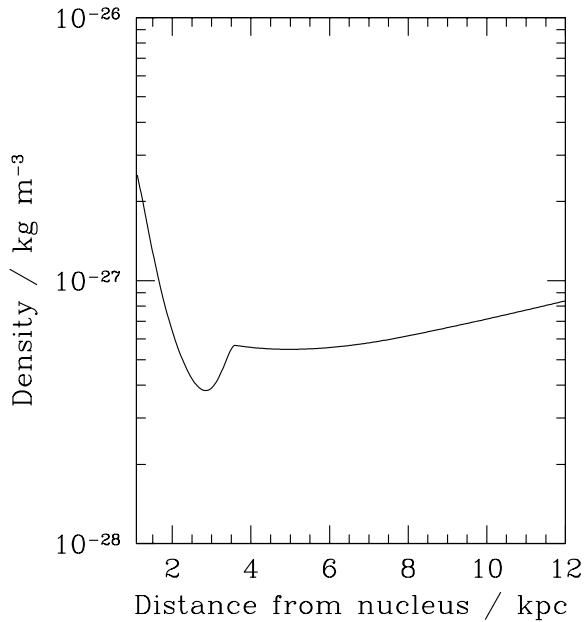}
\caption{The internal density $\rho$ for the reference model.
\label{density-ref}}
\end{figure}

\begin{figure}
\epsfxsize=8.5cm
\epsffile{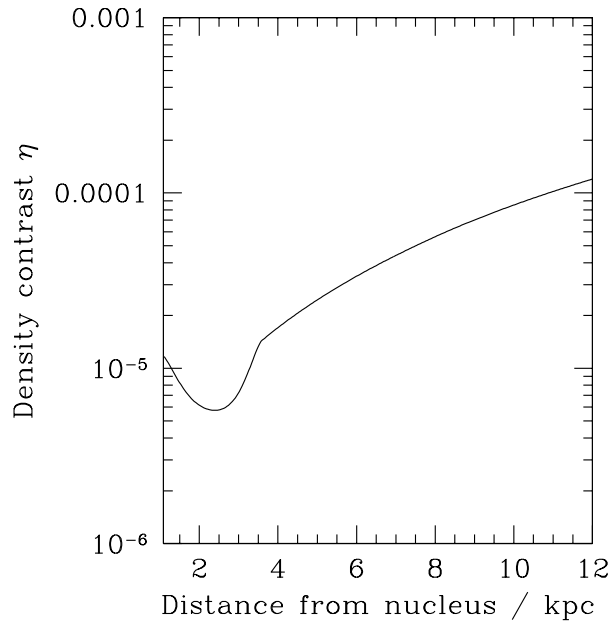}
\caption{The density contrast $\eta = \rho / \rho_{\rm ext}$ for the reference model.
\label{eta-ref}}
\end{figure}

\begin{figure}
\epsfxsize=8.5cm
\epsffile{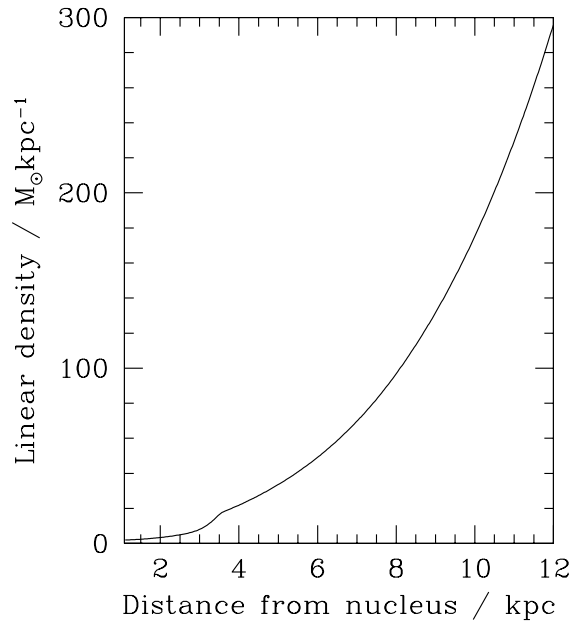}
\caption{The mass per unit length, $\rho A$, for the reference model,
plotted for comparison with the results of \citet{RHCJ99}.
\label{lindens-ref}}
\end{figure}

The internal density (Fig.~\ref{density-ref}) is very low everywhere in the
modelled region and the density contrast $\eta = \rho / \rho_{\rm ext}$ is in
the range $6 \times 10^{-6}$ -- 10$^{-4}$ (Fig.~\ref{eta-ref}).  The density
drops rapidly between the flaring point and 2.7\,kpc from the nucleus (i.e. in
the region of fastest expansion).  Thereafter the jet decelerates abruptly and
begins to recollimate, causing $\rho$ to rise again until the end of the flaring
region. The density remains approximately constant over the outer region despite the
continuing expansion.  Figure~\ref{lindens-ref} plots the mass per unit length,
$\rho A$, for comparison with the ``linear density'' estimated by
\citet{RHCJ99} from numerical simulations of entraining, non-relativistic jets
(Section~\ref{numsym}).

We have also estimated the Faraday rotation expected for a minimum-pressure
field.  For a fully ordered field, the rotation measures would be $\approx$0.03
and $\approx$0.04\,rad\,m$^{-2}$ at the flaring point and at the end of the
modelled region, respectively.  A disordered, but anisotropic field (as assumed
in the models of \citealt{LB02}) would give even lower values. Such low rotation
measures (or any associated depolarization) would be impossible to measure with
the frequency and resolution combinations available to current synthesis arrays,
especially in the presence of significant foreground Faraday rotation. In
3C\,31, the Faraday rotation observed along the jets appears to be produced by a
foreground magnetoionic medium, almost certainly the surrounding hot gas, and
shows fluctuations in the range $-$120 -- +20\,rad\,m$^{-2}$ (Laing et al., in
preparation). There is no evidence for any component internal to the jet.

\subsubsection{Mach number}

\begin{figure}
\epsfxsize=8.5cm
\epsffile{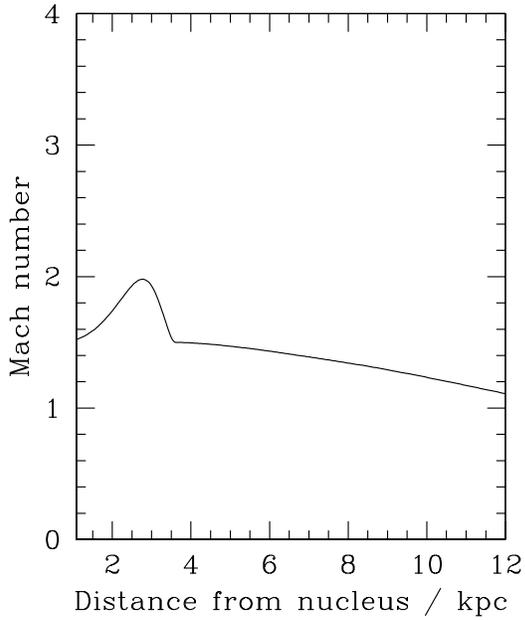}
\caption{The generalized Mach number $\cal M$ for the reference model.
\label{Mach-ref}}
\end{figure}

The flow is always transonic (Fig.~\ref{Mach-ref}), the relativistic Mach
number $\cal M$ reaching a maximum of 2.0 in the flaring region and
falling from 1.5 to 1.1 in the outer region. As pointed out by
\citet{Bic94}, velocities $\sim$0.3 -- 0.7$c$ inevitably correspond to
transonic flow for light, decelerating relativistic jets, and this is
indeed the case for our models.

\subsubsection{Mass flux and entrainment rate}
\label{ent-rate}

\begin{figure}
\epsfxsize=8.5cm
\epsffile{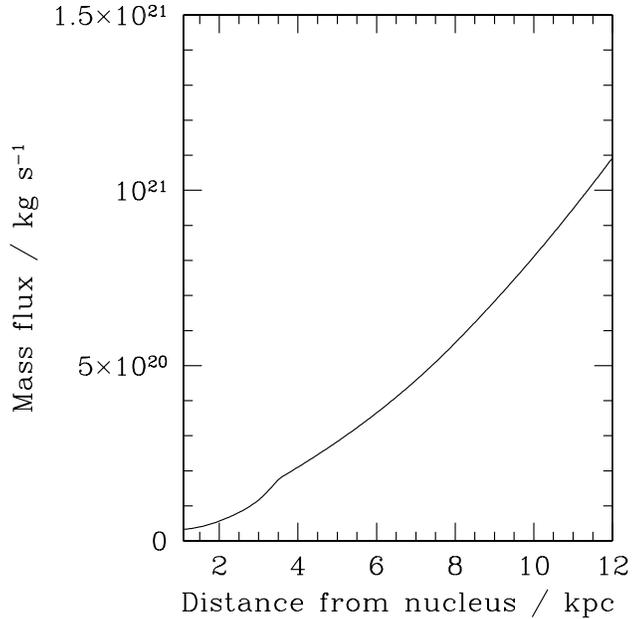}
\caption{The estimated mass flux along the jet for the reference model.
\label{mf-ref}}
\end{figure}

\begin{figure}
\epsfxsize=8.5cm
\epsffile{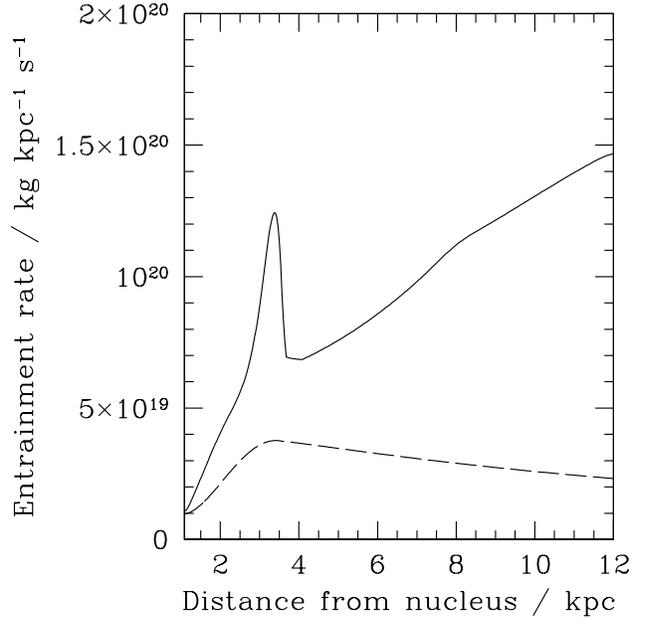}
\caption{The estimated internal mass input rate from stars (long dashes) 
superimposed on the entrainment rate required by the reference model 
 (full line).\label{ent-ref}}
\end{figure}

\begin{figure}
\epsfxsize=8.5cm
\epsffile{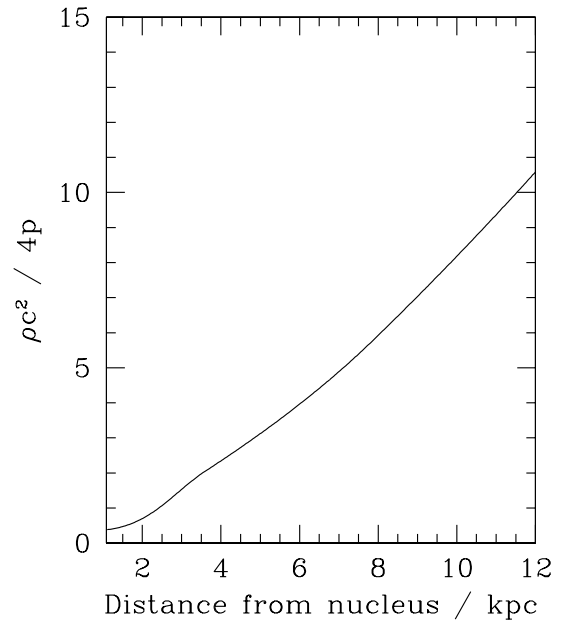}
\caption{\citet{Bic94}'s parameter ${\cal R} = \rho c^2/4p$, which
quantifies the ratio of bulk kinetic to internal energy, for the reference
model. 
\label{r-ref}}
\end{figure}

The mass flux along the jet is plotted in Fig.~\ref{mf-ref}.  Its derivative,
the entrainment rate (the full line in Fig.~\ref{ent-ref}) increases rapidly to
a maximum of 1.3 $\times$ 10$^{20}$ kg\,kpc$^{-1}$\,s$^{-1}$ at a distance of
3.4\,kpc, where the mass flux curve has a point of inflection.  This maximum is
a feature of all acceptable models and is a direct consequence of the rapid
expansion of the jet at roughly constant velocity, followed by abrupt
deceleration. Thereafter, the entrainment rate drops abruptly until the end of
the flaring region and then increases monotonically through the outer region.
The change of slope at 8\,kpc is caused by the discontinuity in the assumed form
for the external temperature gradient.
 
We have independently estimated the rate of entrainment into the jet from stellar mass 
loss, as follows:
\begin{enumerate}
\item We started with the R-band CCD photometry of \citet{Owe89}, which is
well fitted by a power-law surface-brightness distribution
\[ \sigma(R)/{\rm mag\,arcsec}^{-2} = 15.53 - 2.5\delta \lg (R/{\rm arcsec}) \]
with $\delta = 1.65$. A galactic extinction of
$A_R = 0.189$ \citep{Sch98} was removed, a K-correction was applied
as in \citet{Owe89} and the fit was converted to absolute magnitude (note
the change of Hubble Constant to $H_0$ = 70\,km\,s$^{-1}$\,Mpc$^{-1}$
from that used by \citealt{Owe89}).
\item The surface-brightness distribution was then deprojected to give the
luminosity density.  For a surface-brightness distribution of the form
\[ \Sigma(R) = \Sigma_0 R^{-\delta} \]
this is 
\[ L(r) = \left[ \frac{\delta\Sigma_0}{\pi}\int^{\pi/2}_0
(\cos\theta)^\delta d\theta \right] r^{-(1+\delta)} \]
which can be converted to solar luminosities assuming an absolute
magnitude of 4.46 for the Sun in the Kron-Cousins R band \citep{AQ,Fernie}.
\item We then took the mass loss rate predicted by \citet{Fab76} for an
elliptical galaxy stellar population as a function of the blue luminosity,
$L_B$, in solar units,
\[ (\dot{M}/M_\odot\,{\rm yr}^{-1}) = 0.015 (L_B/ 10^9 L_{B\odot}) \]
which is consistent with the estimate from infrared observations by
\citet{Kna92}, and scaled it to the R band using extinction-corrected
colours for NGC\,383 \citep{Sandage,Sch98} and the Sun \citep{AQ}, getting
\begin{equation}
(\dot{M}/M_\odot\,{\rm yr}^{-1}) = 0.0077 (L_R/ 10^9 L_{R\odot})
\label{mdot-eq} 
\end{equation}
\end{enumerate}
This allowed us to estimate the mass-loss rate per unit volume as a function of
radius for 3C\,31.  The assumptions are identical to those of \citet{Kom94},
except for our use of actual galaxy photometry.  The uncertainties are large:
efficient mixing is assumed, and the mass-loss rate is supposed to be completely
unaffected by the presence of a jet.  The dashed line in Fig.~\ref{ent-ref}
shows the estimated entrainment rate from stars, for comparison with that
required to slow the jet according to our analysis.

\begin{figure*}
\epsfxsize=14cm
\epsffile{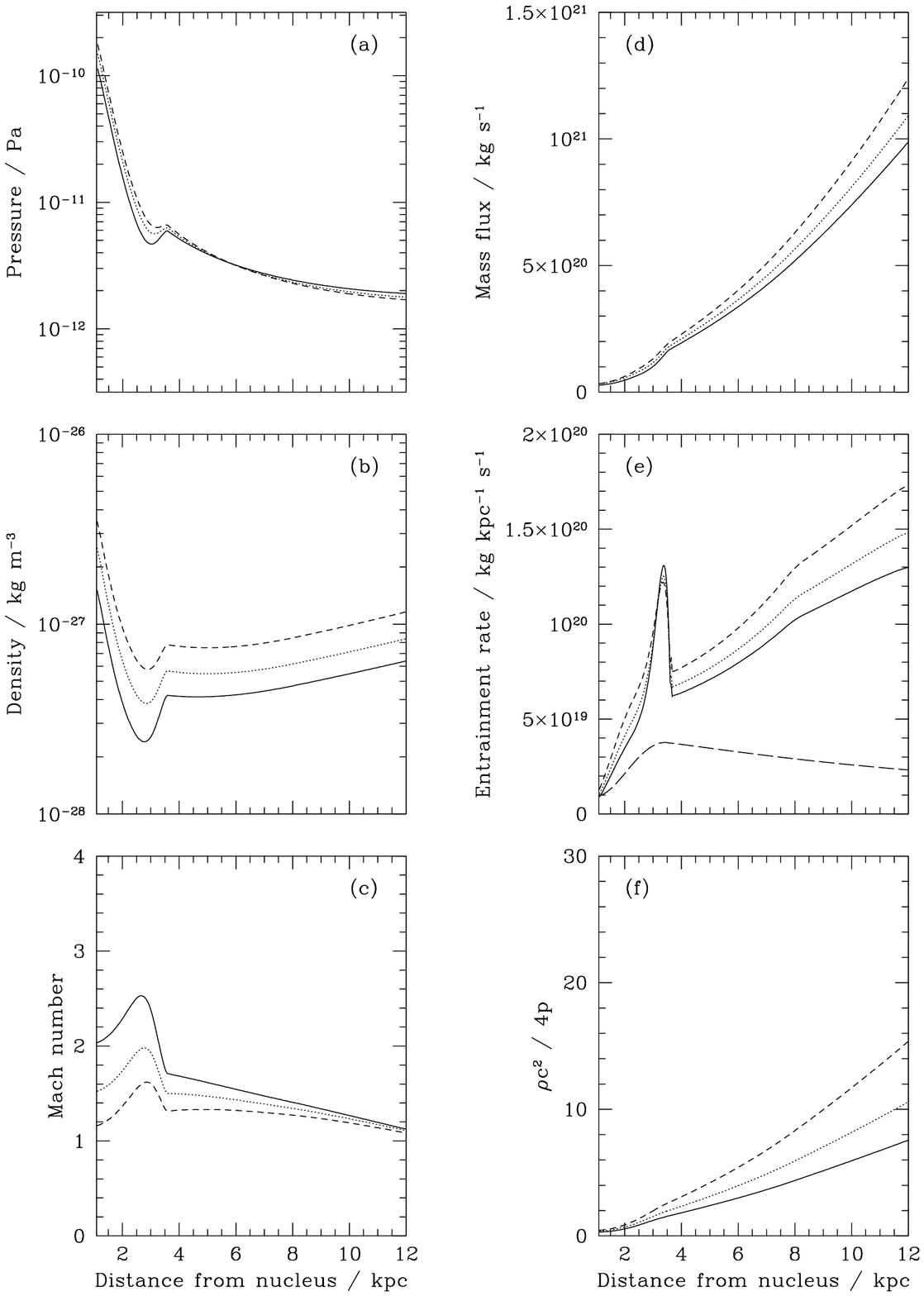}
\caption{Internal flow variables derived from the conservation-law
analysis, as described in the text. The models are pressure-matched in the
outer region and the three curves in each panel represent constant
fractional velocities with $f = 1.0$ (full), 0.85 (dotted) and 0.7 (short
dashed) times the central value. The corresponding energy fluxes are 1.4
$\times$ 10$^{37}$, 1.1 $\times$ 10$^{37}$ and 8.7 $\times$ 10$^{36}$\,W.
(a) pressure; (b) density; (c) Mach number; (d) mass flux; (e) entrainment
rate, with the estimate for stellar mass loss shown by the long dashed
curve; (f) ${\cal R} = \rho c^2 / 4p$. \label{consfit}}
\end{figure*}
 
The required entrainment rate is very close to that expected from stellar mass
loss at the flaring point and remains only a factor of 2 larger until
$\approx$2\,kpc from the nucleus.  Both curves have their maxima at roughly the
same location (close to the point of inflection of the outer isophote), although
that for stellar mass input is much shallower.  It is of course inevitable that
the latter will have a maximum, since the jet expands and the stellar density
decreases with distance from the nucleus. Given the uncertainties, the agreement
between the two estimates in the first 2\,kpc of the flaring region is
surprisingly good. The discrepancy in the outer region is much more significant
as the required injection rate continues to increase while that from stars falls
off.

\subsubsection{Ratio of internal to kinetic energy}

\citet{Bic94} defined the parameter ${\cal R} = \rho c^2/4p$, which
quantifies the ratio of bulk kinetic to internal energy. This
increases monotonically from an initial value of 0.4 to 10.6 at the end
of the modelled region (Fig.~\ref{r-ref}). 

\subsection{Other pressure-matched solutions}
\label{vel-var-peq}

For constant-fraction velocity profiles, there are physically self-consistent,
pressure-matched solutions for any velocity fraction $f$ in the range $0.6 \leq
f \leq 1$.  For $f < 0.6$, pressure equilibrium in the outer region is
impossible. This value of $f$ is in any case slightly lower than the minimum
(edge) velocity, $f = 0.7$, in our kinematic model \citep{LB02}, which we take
as a lower limit in what follows.  As noted in Section~\ref{refmod}, the case $f
= 1$, in which all of the material travels at the on-axis velocity, is unlikely
to be approached in practice, but we retain it as an upper bound on acceptable
solutions.  Fig.~\ref{consfit} shows the variation of the flow variables with
distance along the jet axis for pressure-matched models with $f = 0.70$
(minimum), $f = 0.85$ (the reference model, as before) and $f = 1$
(on-axis). The internal pressures are very similar by construction, but the main
qualitative features of the reference model also remain for the other flow
variables: there is a density minimum in the flaring region, corresponding to a
maximum Mach number, and a local peak in the entrainment rate. Faster jets
(larger $f$) have higher energy fluxes and Mach numbers, but lower densities and
entrainment rates.  We found an accurate empirical linear relation between $\log
\Phi$ and $f$ from results for 9 values of $f$ in the range $0.6 \leq f \leq 1$.
The coefficients, determined by a least-squares fit, are given in
Table~\ref{results-table}.  The energy flux is constrained to better than a
factor of 2, varying from $8.7 \times 10^{36}$\,W for $f = 0.7$ to $1.4 \times
10^{37}$\,W for $f = 1$.

As expected, the pressure-matched solutions for the ramped velocity
profile are very similar to those for a fixed fractional velocity and $f =
1$ (Fig.~\ref{consfit}) and are therefore not shown.  There is a similar,
but flatter, relation between $\log \Phi$ and $f$
(Table~\ref{results-table}) and the energy flux lies in a very narrow
range from $1.1 \times 10^{37}$\,W ($f = 0.7$) to $1.4 \times 10^{37}$\,W
($f = 1$).  Other velocity laws do not require significantly different
energy fluxes or flow variables.

Given the uncertainties inherent in the quasi-one-dimensional
approximation, Fig.~\ref{consfit} therefore represents our best estimate
of the range of allowed solutions subject to the reasonable assumptions that  
the outer jet is in pressure balance and that $\Pi = \Phi/c$.

\subsection{Review of the assumptions}
\label{vars}

\subsubsection{Uncertainties in the on-axis velocity model}
\label{vel-errors}

\begin{figure*}
\epsfxsize=14cm
\epsffile{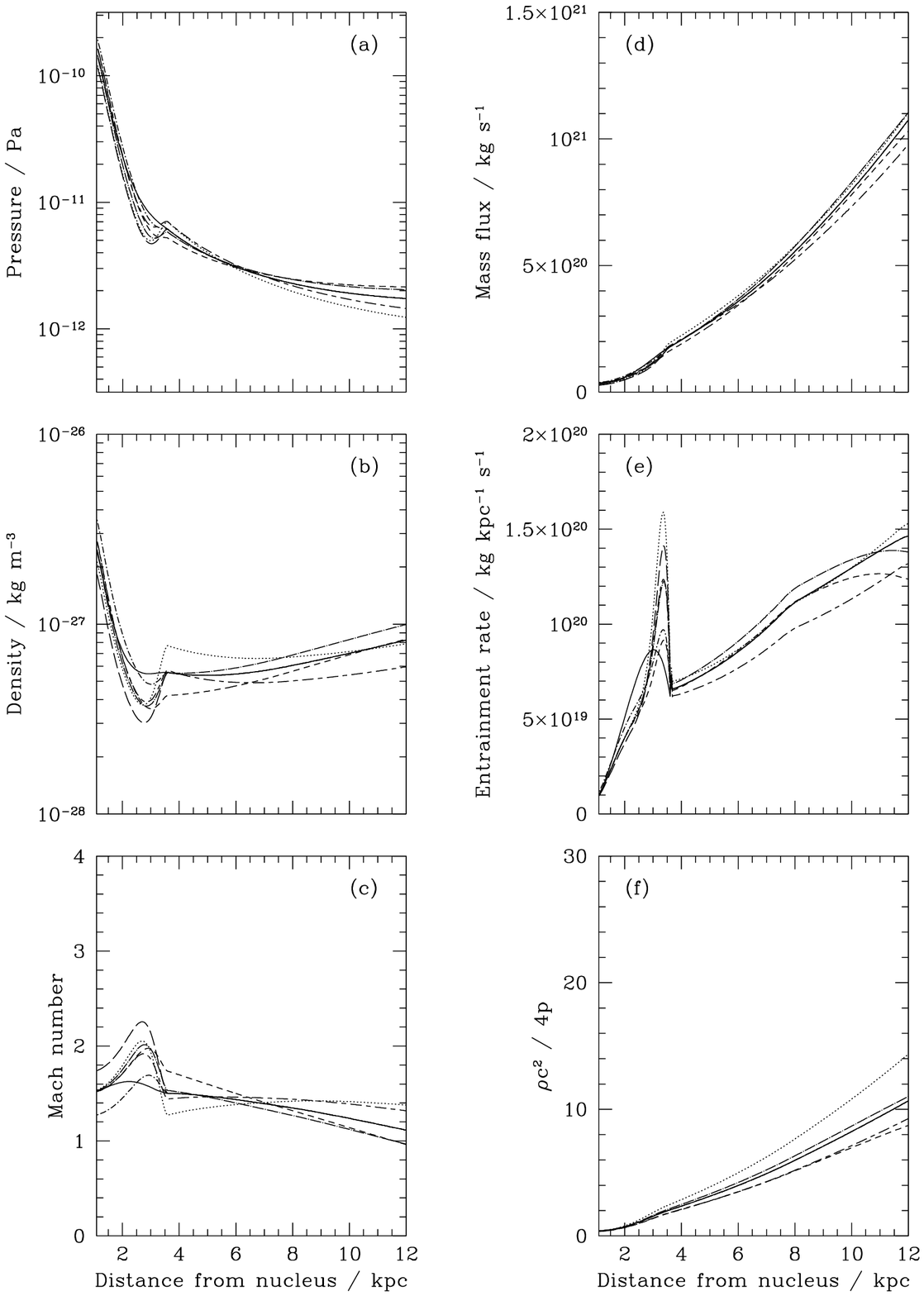}
\caption{Curves showing the effects of the uncertainties in the
velocity field on the values of the derived flow variables.
Seven curves are overlaid for each of the flow variables
shown in Fig.~\ref{consfit}, for a fractional
velocity $f=0.85$ and pressure-matching in the outer region.
The curves result from setting the
fiducial velocities and exponent (Fig.~\ref{vel-profile}) in turn to the
maxima and minima given in table 7 of \citet{LB02}, 
leaving the other parameters at their nominal
values.  The curves are: $H = 3.6$ (full); $\beta_1
= 0.83$ (long dash); $\beta_1 = 0.68$ (dot -- short dash); $\beta_0 = 0.45$
(dotted); $\beta_0 = 0.63$ (short dash); $\beta_f = 0.25$ (dot -- long dash) and
$\beta_f = 0.33$ (short dash -- long dash).  (a) pressure; (b) density; (c) Mach
number; (d) mass flux; (e) entrainment rate; (f) ${\cal R} =
\rho c^2 / 4p$.
\label{vel-err-fig}}
\end{figure*}

Uncertainties in fitting the kinematic model to the observed synchrotron
emission result in errors in the on-axis velocity.  We have investigated the
effect of varying the velocity exponent and each of the three fiducial
velocities (Fig.~\ref{vel-profile}) independently over the allowed ranges given
in table 7 of \citet{LB02}.  Fig~\ref{vel-err-fig} superposes seven curves to
illustrate how the on-axis velocity uncertainties affect the flow parameters
inferred for the reference model of Section~\ref{refmod} (the pressure-matched
solution with constant velocity fraction $f = 0.85$ and $\Phi = \Pi c$).  The
energy flux is barely affected (9 $\times$ 10$^{36}$\,W $\leq \Phi \leq$ 1.2
$\times$ 10$^{37}$\,W).  The most significant changes are in the density and
peak entrainment rate, where the maximum spread approaches a factor of 2. The
uncertainties are generally smaller than those introduced by varying the
velocity fraction, $f$ (Fig.~\ref{consfit}).  One feature of interest is that
reducing the velocity exponent $H$ \citep{LB02} to its lowest allowed value,
thereby smoothing out the abrupt deceleration at the end of the flaring region,
also reduces the amplitude of the fluctuations of pressure, density, Mach number
and entrainment rate.

\subsubsection{External gas density and pressure}
\label{ext-var}

\begin{figure*}
\epsfxsize=14cm
\epsffile{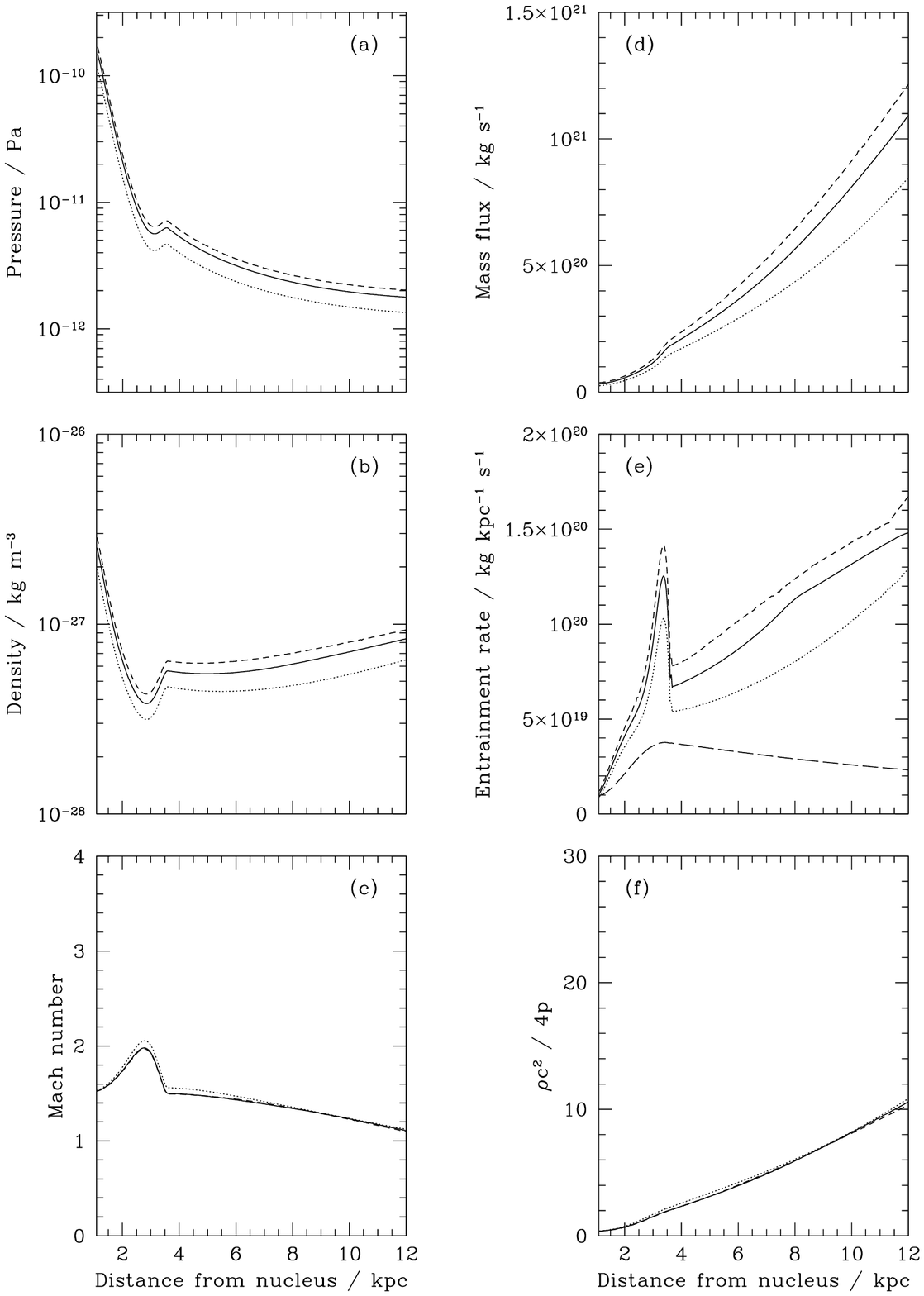}
\caption{Curves showing the effects of uncertainties in the external pressure
and temperature on the flow variables derived from the conservation-law
analysis. The models are pressure-matched in the outer region and have constant
velocity fraction $f = 0.85$.  The three curves in each panel demonstrate the
effects of altering the external density and pressure by the combined $1\sigma$
errors in the fitting procedure from the nominal profiles \citep{Hard_3C31}.
The curves are: nominal (full line), increased pressure (short dashed) and
decreased pressure (dotted).  The corresponding energy fluxes are 1.1 $\times$
10$^{37}$, 1.3 $\times$ 10$^{37}$ and 8.5 $\times$ 10$^{36}$\,W.  (a) pressure;
(b) density; (c) Mach number; (d) mass flux; (e) entrainment rate, with the
estimate for stellar mass loss shown by the long dashed curve; (f) ${\cal R} =
\rho c^2 / 4p$.\label{consfitd}}
\end{figure*}

In order to quantify the uncertainties introduced by errors in fitting the
{\em Chandra} data, we have repeated the analysis of Section~\ref{refmod}
with density and pressure profiles corresponding to combined $\pm 1\sigma$
uncertainties in the conversion between central normalization and density,
the $\beta$ model fits and the linear fit to the temperature gradient (as
in fig.\,9 of \citealt{Hard_3C31}).  The changes to our solutions are
small compared with those due to other uncertainties.  The energy flux for
a pressure-matched solution changes by $\pm$20\% and the derived profiles
of flow variables are not grossly affected, as illustrated for a constant
velocity fraction $f = 0.85$ and $\Phi = \Pi c$ in Fig.~\ref{consfitd}.
The pressure, density and entrainment rate are typically \la 25\% from the
reference model.  We conclude that errors in the fitting of the
external gas parameters are not serious compared with other uncertainties
in the problem.

\subsubsection{Solutions which are not pressure-matched in the outer region}
\label{no-p-balance}

\begin{figure*}
\epsfxsize=14cm
\epsffile{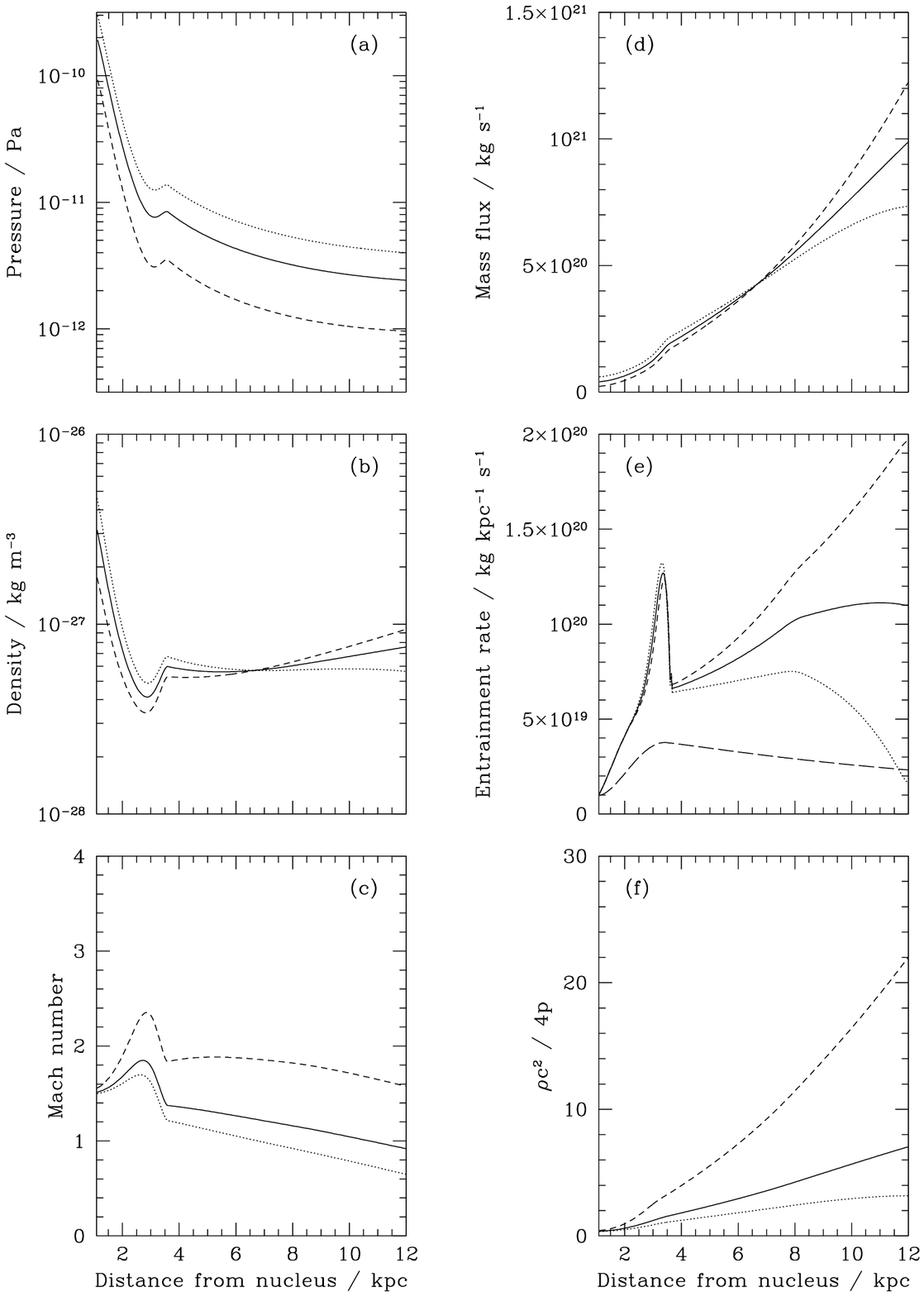}
\caption{Curves showing the effects of relaxing the requirement for pressure
matching in the outer region. The flow variables are derived from the
conservation-law analysis for a fractional velocity $f = 0.85$.  The three
curves in each panel represent the minimum energy flux allowed by the
synchrotron pressure condition (short dashed), the maximum consistent with an
instantaneous over-pressure of a factor of 10 (full) and the upper limit set by
the requirement that the mass flux increases monotonically away from the nucleus
(dotted).  (a) pressure; (b) density; (c) Mach number; (d) mass flux; (e)
entrainment rate, with the estimate for stellar mass loss shown by the long
dashed curve; (f) ${\cal R} = \rho c^2 / 4p$.
\label{consfita}}
\end{figure*}

The effect of relaxing the assumption of pressure balance in the outer region is
illustrated in Fig.~\ref{consfita}. This shows the same plots as in
Fig.~\ref{consfit} for a constant value of $f = 0.85$, but a range of energy
fluxes corresponding to the overpressure and mass-flux conditions (ii), (iii)
and (v) given in Section~\ref{accept}.  Here, the internal {\em densities} are
similar, but pressure, Mach number, mass flux and entrainment rate change
significantly.  In each case, the {\em lower} limit to the energy flux is set by
the condition that the internal pressure must exceed the synchrotron minimum
pressure.  The absolute {\em upper} limit is determined by the condition that
the mass flux must always increase along the jet, although values towards the
upper end of the range require very large over-pressures in the flaring region,
and we have flagged those solutions which have $p > 10p_{\rm ext}$ at any point.

The energy flux ranges of acceptable solutions for the velocity laws
assuming constant fractions of the central speed are shown in
Fig.~\ref{soln-parms}.  At a given value of $f$, the acceptable range is
roughly a factor of 2. The energy flux
is, of course, an increasing function of the velocity fraction.  Again,
the relations between $f$ and $\log \Phi$ are accurately linear for the
limiting cases, and we have plotted the best fits in Fig.~\ref{soln-parms}
and summarized the results in Table~\ref{results-table}.  Relaxation of
the pressure balance condition increases the allowed energy flux range:
the minimum value is $4.5 \times 10^{36}$\,W ($p = p_{\rm sync}$; $f =
0.7$); the maximum is either $2.3 \times 10^{37}$\,W (overpressure
condition; $f = 1$) or $3.0 \times 10^{37}$\,W (mass flux condition; $f =
1$).

The solutions that are significantly over-pressured in the outer region
have relatively low entrainment rates which drop at large distances from
the nucleus (Fig.~\ref{consfita}e) and which are therefore closer to
the estimated mass input from stars.  At first sight,
these could represent free expansion without external entrainment,
but they have low Mach numbers ${\cal M} \approx 1$
(Fig.~\ref{consfita}c), so the observed opening angle is not consistent
with this idea.
 
\begin{table}
\caption{Expressions giving the energy fluxes for pressure-matched and
limiting solutions as functions of the fraction velocity $f$:~ 
$\log (\Phi / {\rm W}) = Af + B$. The relations are accurate to $\approx$0.01
rms in $\log \Phi$. \label{results-table}}
\begin{tabular}{lrr}
\hline
&&\\
Solution description & \multicolumn{1}{c}{A} & \multicolumn{1}{c}{B} \\
&&\\
\multicolumn{3}{c}{Fixed fractional velocity}\\
&&\\
Pressure-matched & 0.777 & 36.38 \\
Lower limit ($p = p_{\rm sync}$) & 1.080 & 36.01 \\
Upper limit ($p = 10 p_{\rm ext}$ in flaring region) & 1.293 & 36.07 \\
Upper limit (monotonically increasing mass flux) & 0.853 & 36.63 \\
&&\\
\multicolumn{3}{c}{Ramped fractional velocity}\\
&&\\
Pressure-matched & 0.335 & 36.84 \\
Lower limit ($p = p_{\rm sync}$) & 0.190 & 36.91 \\
Upper limit (monotonically increasing mass flux) & 1.040 & 36.45 \\
&&\\
\hline
\end{tabular}
\end{table}

\subsubsection{Synchrotron minimum pressure}
\label{sync-press-var}

Changes in the assumptions made in deriving the synchrotron minimum pressure
affect the range of allowed solutions by influencing the lower energy flux bound
(Fig.~\ref{soln-parms} and Table~\ref{results-table}).  The only change which
can reduce $p_{\rm sync}$ is to curtail the energy range of the relativistic
particles.  So far, we have assumed a range from $\gamma_{\rm min} = 10$ to
$\gamma_{\rm max} = 10^5$.  If we increase $\gamma_{\rm min}$ to 1000, $p_{\rm
sync}$ decreases by a factor of 0.59 (equation~\ref{eq-pmin}).  This would
reduce the minimum allowed energy flux from $8.7 \times 10^{36}$\,W to $5.8
\times 10^{36}$\,W, but in fact the under-pressure condition in the outer region
takes over to set the minimum energy flux at $6.9 \times 10^{36}$\,W. The
derived profiles are not changed greatly and the pressure-matched solution is
unaffected.  Even if we take an unrealistically small range of $\gamma$ from
10$^3$ -- 10$^4$ (equivalent to a factor of 100 in frequency, which would
preclude an extension of the spectrum into the X-ray band), the decrease in
$p_{\rm sync}$ is only by a factor of 0.42.  There is, of course, no reason to
suppose that the energy limits are independent of position.

Other changes (increasing the energy range, decreasing the filling factor,
changing the ratio between field and particle energy or adding relativistic
protons) act to increase the pressure. This will tend to reduce the range of
solutions, eventually causing a conflict with the pressure-matching and then the
over-pressure conditions.  At present, however, the inferred internal, external
and synchrotron minimum pressures are mutually consistent.

\subsubsection{Equation of state}
\label{eos}

The fact that $p$ is only a factor of $\approx$2 larger than $p_{\rm sync}$ in
the outer region is consistent with our initial assumption that relativistic
particles and magnetic field dominate the internal energy of the jet, and gives
a post hoc justification for the adoption of an ultra-relativistic equation of
state.  If there are significant departures from equipartition, then $p_{\rm
sync}$ will be significantly higher, strengthening our conclusion.  Even if we
adopt the most conservative limits on the particle energy spectrum
(Section~\ref{sync-press-var}), relativistic particles and field must contribute
a large fraction of the internal energy of the jet material.

The density of the entrained material remains much lower than that of the
external medium ($\eta \approx 10^{-4}$ at the end of the modelled
region), justifying the assumption that the entrained  internal energy
is negligible compared with that of the relativistic particles.
The ultra-relativistic equation of state should therefore be an adequate
approximation for our purposes, but will not necessarily remain valid far
from the nucleus.

A full treatment using the equation of state for mixed relativistic and
non-relativistic plasma \citep{Synge57,Kom94} is outside the
scope of this paper.  We have, however, tested the sensitivity of our
results to the choice of equation of state by computing the (physically
unrealistic) limiting case of a pure non-relativistic plasma.  We find
that the solutions are similar to those for the ultrarelativistic equation
of state. For the reference model, the energy flux ($\Phi = 7.2 \times
10^{36}$\,W) is slightly smaller;  the internal pressure is essentially
unchanged, and the density, mass flux and entrainment rate in the outer
region are lower by at most 25\%.  The only major differences occur close
to the flaring point, where the solutions for a non-relativistic equation
of state give densities and entrainment rates lower by factors of $\sim$4
and $\sim$2, respectively.

\subsubsection{Momentum flux}
\label{mom-flux-var}

\begin{figure*}
\epsfxsize=14cm
\epsffile{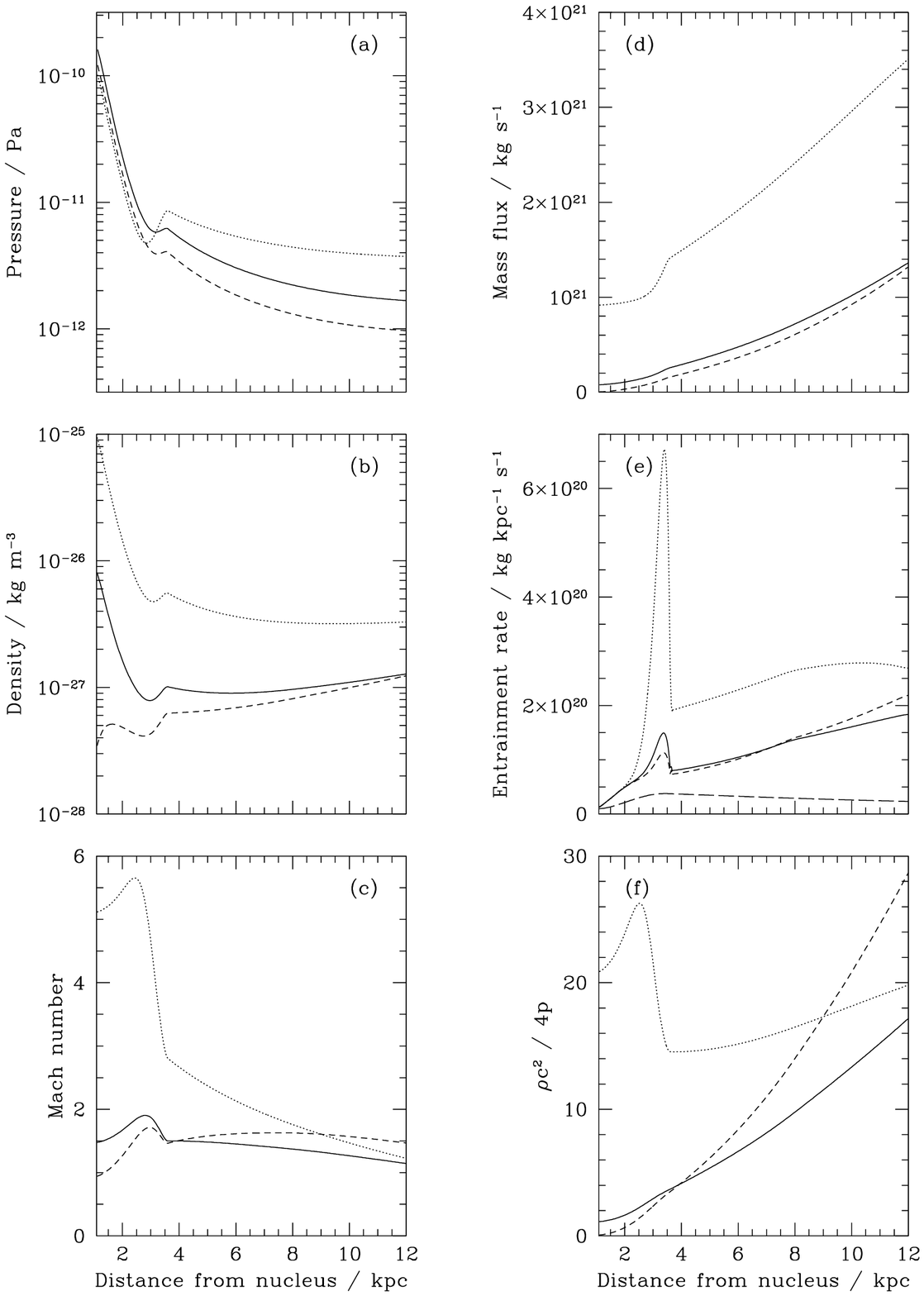}
\caption{Curves showing the effects of relaxing the condition $\Phi = \Pi c$.
The flow variables were derived from the conservation-law analysis for a
fractional velocity $f = 0.7$.  The three curves in each diagram represent: the
maximum momentum and energy fluxes (dotted lines), an energy flux of 8.7
$\times$ 10$^{36}$\,W, as in Fig.~\ref{consfit}, but with a momentum flux
corresponding to the asymptotic value given by equation~(\ref{asymp-R}) for
$\Gamma = 5$ (full lines) and the minimum energy and momentum fluxes (short
dashed lines).  (a) pressure; (b) density; (c) Mach number; (d) mass flux; (e)
entrainment rate, with the estimate for stellar mass loss shown by the long
dashed curve; (f) ${\cal R} = \rho c^2 / 4p$.  Note that the ranges for density,
mass flux, Mach number and entrainment rate are larger than those in
Figs~\ref{consfit}, \ref{vel-err-fig}, \ref{consfitd} and \ref{consfita}.
\label{consfitb}}
\end{figure*}

We now explore the effect of relaxing the assumption that $\Phi = \Pi c$.
Unless the velocity is very close to $c$ on parsec scales, the momentum flux
$\Pi$ is potentially a free parameter, depending on the initial value of ${\cal
R}$, the ratio of rest mass energy to enthalpy (equation~\ref{R-eqn}).  We have
searched for acceptable solutions (according to the criteria (i) -- (vi) of
Section~\ref{accept}) over a grid of values of $\Phi$ and $\Pi$ for models with
fixed fractional velocities $f =$ 0.7, 0.85 and 1.0.  As expected, there are no
solutions with $\Pi$ significantly less than $\Phi/c$, but there are solutions
with high momentum flux.  The largest range of solutions is found for fixed
fractional velocity and $f = 0.7$.  We show the variation of flow variables with
distance for two extreme cases (minimum and maximum $\Phi$ and $\Pi$) in
Fig.~\ref{consfitb}. The high-momentum-flux solutions (e.g. that represented by
the dotted curves in Fig.~\ref{consfitb}) are very different from those with
$\Pi = \Phi/c$:

\begin{figure}
\epsfxsize=8.5cm
\epsffile{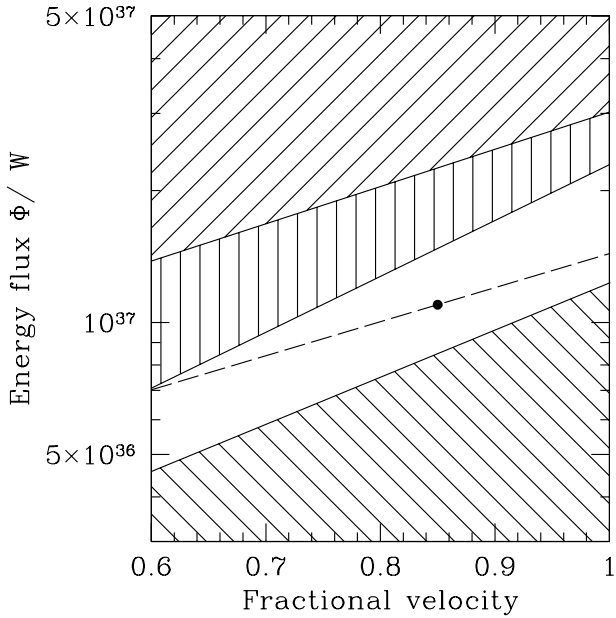}
\caption{A plot of energy flux against fractional velocity showing the
constraints for velocities which are fixed fractions $f$ of the on-axis
value.  Solutions are allowed in the unhatched area at the centre of the
diagram and the dashed line represents the pressure-matched case.  The
filled circle denotes the reference model of Section~\ref{refmod}. The
lower hatched area is excluded by the condition $p > p_{\rm sync}$.
Vertical hatching corresponds to jets with $p > 10 p_{\rm ext}$ at some
point. Jets in the upper, diagonally-hatched region have mass fluxes which
decrease with distance from the nucleus.\label{soln-parms}}
\end{figure}

\begin{figure}
\epsfxsize=8.5cm
\epsffile{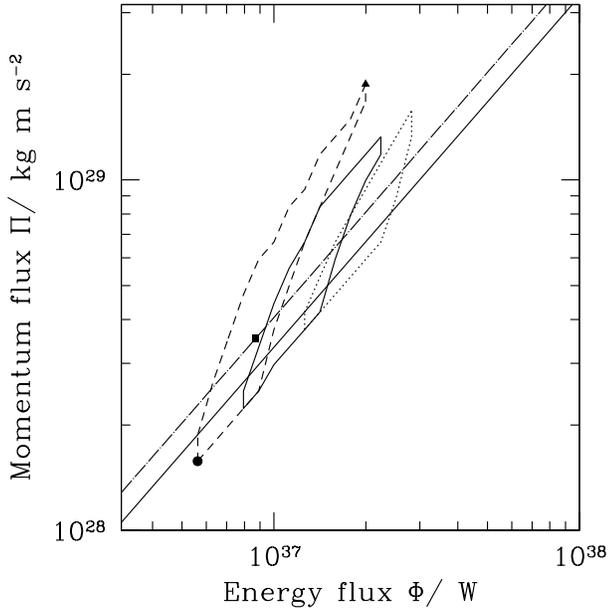}
\caption{A plot of momentum flux against energy flux, showing the allowed
solution ranges for constant-fraction velocities $f$ = 0.70 (dashed),
0.85 (full) and 1.0 (dotted). Solutions can be found inside the closed
figures.  The three solutions for $f = 0.7$ plotted in Fig.~\ref{consfitb}
are also indicated: minimum $(\Phi,\Pi)$ (circle), asymptotic from
equation~(\ref{asymp-R}) for $\Gamma = 5$ (square) and maximum $(\Phi,\Pi)$
(triangle). The full line represents $\Phi = \Pi c$ and the dash-dot line
$\Phi = 0.82\Pi c$, representing the asymptotic ratio for $\Gamma = 5$, as
described in the text.  Solutions above and to the left of this line could
not have decelerated from $\Gamma = 5$ on parsec scales.
\label{mom-var-fig}}
\end{figure}

\begin{enumerate}
\item The densities and values of $\cal R$ are high initially and remain so.
\item Consequently, the mass flux and the entrainment rate are greatly
increased, and the maximum in the entrainment rate in the flaring region
is enhanced.
\item The jets are supersonic in the flaring region,  becoming
transonic only in the outer region.
\end{enumerate}
The range of solutions is plotted in Fig.~\ref{mom-var-fig} for $f =$ 0.7, 0.85
and 1.0.  As expected, the lower bound is just below the line $\Phi = \Pi c$.
The upper bound is set by the condition that the outer region is over-pressured
by less than a factor of 2.  The positions of the solutions closest to pressure
balance in the outer region are also indicated.  In some cases, these have
momentum fluxes significantly in excess of $\Phi/c$, but in no case is the mean
pressure ratio significantly different from that for the equivalent
pressure-matched solution with $\Pi = \Phi/c$.  We have also found solutions for
ramped velocity profiles. The allowed ranges of $\Phi$ and $\Pi$ overlap the
comparable areas for constant-fraction velocity laws, but are considerably
smaller.

\begin{figure}
\epsfxsize=7cm
\epsffile{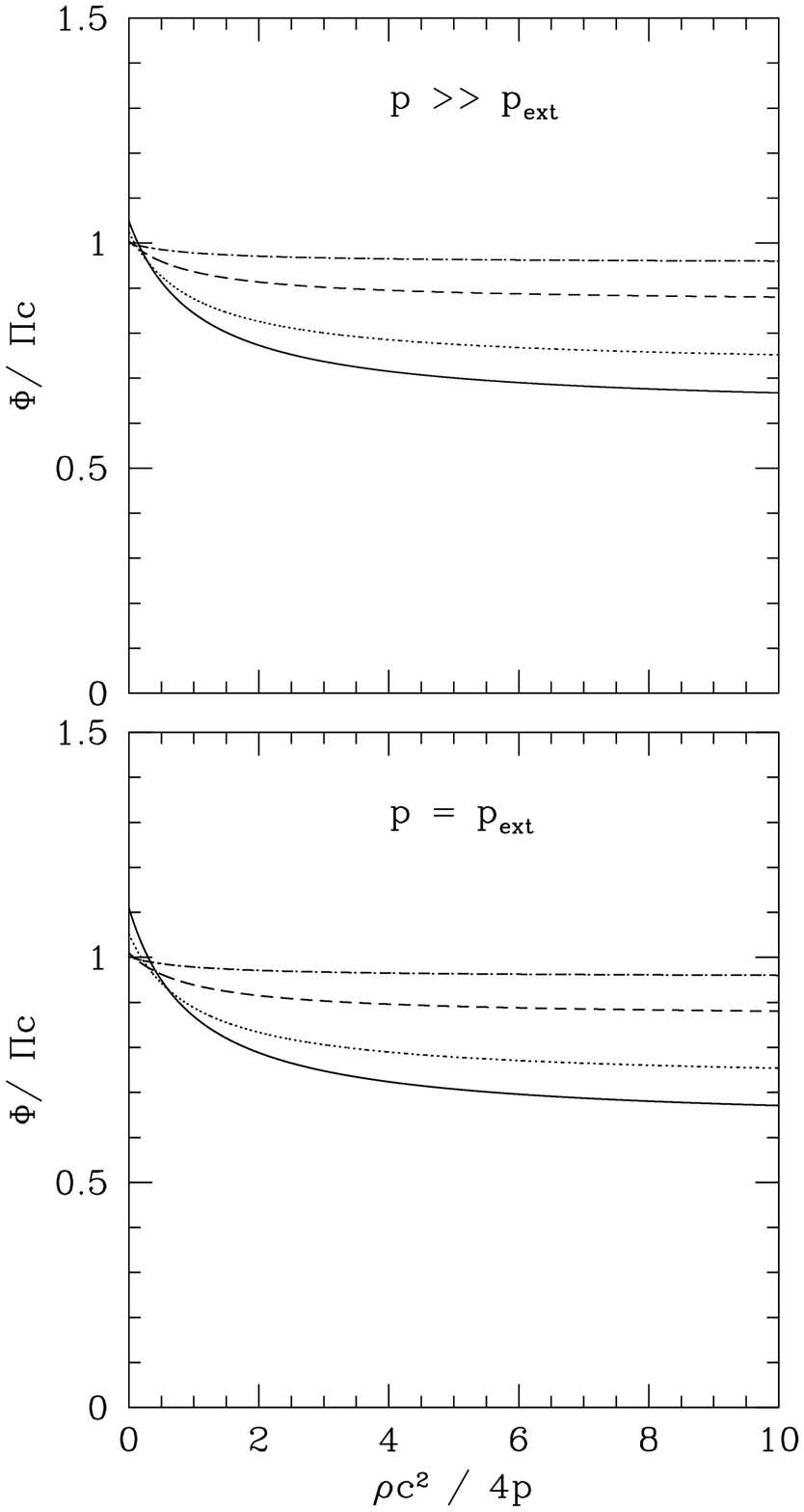}
\caption{Plots of the energy/momentum flux ratio $\Phi/\Pi c$ against
${\cal R } = \rho c^2/4p$ in the absence of buoyancy. Top panel:
over-pressured jet ($p \gg p_{\rm ext}$); bottom panel: pressure
equilibrium ($p = p_{\rm ext}$). The curves correspond to different
velocities: $\beta =$ 0.9 (full line), 0.95 (dots), 0.99 (short dashes) and
0.999 (dash-dots).
\label{momratio-R-fig}}
\end{figure}

There is therefore a set of allowed solutions where the jets are much
heavier than those we have described previously, and so require much more
entrainment in order to slow down.  They are, however, inconsistent with
the deceleration from high Lorentz factors on parsec scales required by
Unified Models.  In the absence of buoyancy, the ratio of energy and
momentum fluxes for a jet in pressure equilibrium ($p = p_{\rm ext}$) can
be written:
\begin{equation}
\frac{\Phi}{\Pi c} = \frac{[(\Gamma^2-\Gamma){\cal R} +
\Gamma^2]\beta}{(\Gamma^2-1)(1+{\cal R})}
\end{equation}
where we have combined equations~(\ref{E-flux}), (\ref{mom-flux}) and (\ref{R-eqn}).
Alternatively, if the jet is very over-pressured ($p \gg p_{\rm ext}$), we
have:
\begin{equation}
\frac{\Phi}{\Pi c} = \frac{[(\Gamma^2-\Gamma){\cal R} +
\Gamma^2]\beta}{(\Gamma^2-1)(1+{\cal R})+1/4}
\end{equation}
These relations are illustrated in Fig.~\ref{momratio-R-fig} as plots of
$\Phi/\Pi c$ against ${\cal R}$ for the cases of pressure equilibrium and
extreme overpressure on parsec scales. 

For a given speed, and regardless of the pressure condition, $\Phi/\Pi c$
must always be larger than the asymptotic value corresponding to ${\cal R}
\rightarrow \infty$:
\begin{equation}
\frac{\Phi}{\Pi c} \geq \frac{(\Gamma^2-\Gamma)\beta}{\Gamma^2-1}
\label{asymp-R}  
\end{equation}
For example, $\Phi/\Pi c \geq
0.71$ if $\Gamma = 3$ and $\Phi/\Pi c \geq 0.82$ if $\Gamma = 5$.  For
light (${\cal R} \ll 1$) and fast ($\beta \rightarrow 1$) jets, $\Phi/\Pi
c$ is very close to 1.  We indicate the asymptotic ratio for $\Gamma = 5$
in Fig.~\ref{mom-var-fig}: it is clear that the majority of high-momentum
solutions are excluded if the Lorentz factors on pc scales are as high as
those required by Unified Models.  We also note that \citet{BB} argue that
${\cal R}$ \la 2 on kpc scales in the M\,87 jet.  We have plotted the flow
variables for a case with $\Phi = 8.7 \times 10^{36}$\,W and $\Phi /\Pi c
= 0.82$ in Fig.~\ref{consfitb}.  The jets in this case are only slightly
denser than for $\Phi = \Pi c$ (compare Fig.~\ref{consfit}).

There are other problems with the very dense solutions: 
\begin{enumerate}
\item They are significantly and persistently over-pressured in the outer
region.
\item The enormous increase in entrainment rate in the flaring region
(Fig~\ref{consfitb}e) occurs despite the fact that the jet is still highly
supersonic (${\cal M} \approx 5$; Fig~\ref{consfitb}c).
\item The ratio $\cal R$, instead of increasing monotonically along the
jet (as expected), has a maximum in the flaring region.
\end{enumerate}
We conclude that while we cannot rule out the high-momentum solutions from our
conservation analysis alone, they would have unphysical properties even on kpc
scales and are inconsistent with deceleration from highly relativistic
speeds on pc scales.  

\section{Discussion}
\label{discussion}

\subsection{General}

Our analysis shows that the hypothesis that the jets decelerate by
entrainment and are recollimated by the external pressure gradient is
quantitatively consistent with our model velocity field for external gas
parameters derived from {\em Chandra} measurements.  The uniqueness of our
solution depends primarily on the assumptions that the jets are in
pressure equilibrium with the external medium in the outer region and that
the momentum flux $\Pi = \Phi/c$.  We have argued that both assumptions
are likely to be correct, but have also demonstrated the effects of
relaxing them.  In the remainder of this Section, we assume that they hold
precisely.

\subsection{Comparison with numerical simulations}
\label{numsym}

Several groups have made numerical hydrodynamic or magnetohydrodynamic
simulations of the effects of internal or external entrainment on jets.
\begin{enumerate}
\item \citet{DeY96} modelled the development of turbulent
eddies and subsequent entrainment, described as ``ingestion'' followed by
``digestion''; 
\item \citet{BLK96} studied the effects of mass input from stars
(considered as a continuous mass source) on a two-dimensional,
relativistic jet;
\item \citet{Lok96} and \citet{Lok97} modelled external entrainment into a
non-relativistic jet using 3D hydrodynamical simulations
and derived a mass entrainment rate.
\item \citet{RHCJ99} and \citet{RH00} investigated the effects of magnetic
fields via 3D, MHD simulations of non-relativistic jets. These showed the
non-linear development of Kelvin-Helmholtz (KH) instabilities and their role in
external entrainment.
\end{enumerate}
Although all of these simulations give important insights into the physics
of jet deceleration, none can be compared directly with our results.
Except for the calculations of \citet{BLK96}, which deal specifically with
mass input from stars and exclude external entrainment, all are
non-relativistic.  Without exception, they assume that the jets are
initially in pressure equilibrium with the external medium.  This is
inconsistent with our inference of a significant over-pressure in the
flaring region.  Only \citet{BLK96} include a realistic galactic
atmosphere.  Finally, the jets in the highest-resolution three-dimensional
simulations \citep{RH00} are much denser than we infer ($\eta = 0.25$,
compared with $\approx$10$^{-5}$).

Nevertheless, comparison of Fig.~\ref{lindens-ref} with fig.\,2 of
\citet{RHCJ99} shows some  similarities: the linear density 
initially grows slowly, then increases rapidly through the flaring region
and levels off at the beginning of the outer region.  The three phases are
interpreted as the linear, non-linear and saturated stages of the
Kelvin-Helmholtz instabilities.  There are obvious differences, however: the
simulations do not show an abrupt increase in emissivity at the flaring
point, nor do they predict the further rapid increase in entrainment rate
in the outer region.  

\subsection{The onset of deceleration}

A common feature of all our acceptable solutions is that the jet becomes
significantly over-pressured at the start of the flaring region.  Such a
localised region can persist in a steady-state jet \citep{Leahy91}: it is
apparently ``unconfined'', but the fluid passing through it is expanding,
and by the time it has expanded, it is further down the jet, and close to
pressure equilibrium with the surroundings.  In fact, the rapid expansion
in this region causes the pressure to drop abruptly and the jet becomes
over-expanded, starts to recollimate and attains pressure equilibrium over
roughly a sound crossing distance.

\begin{figure}
\epsfxsize=8.5cm
\epsffile{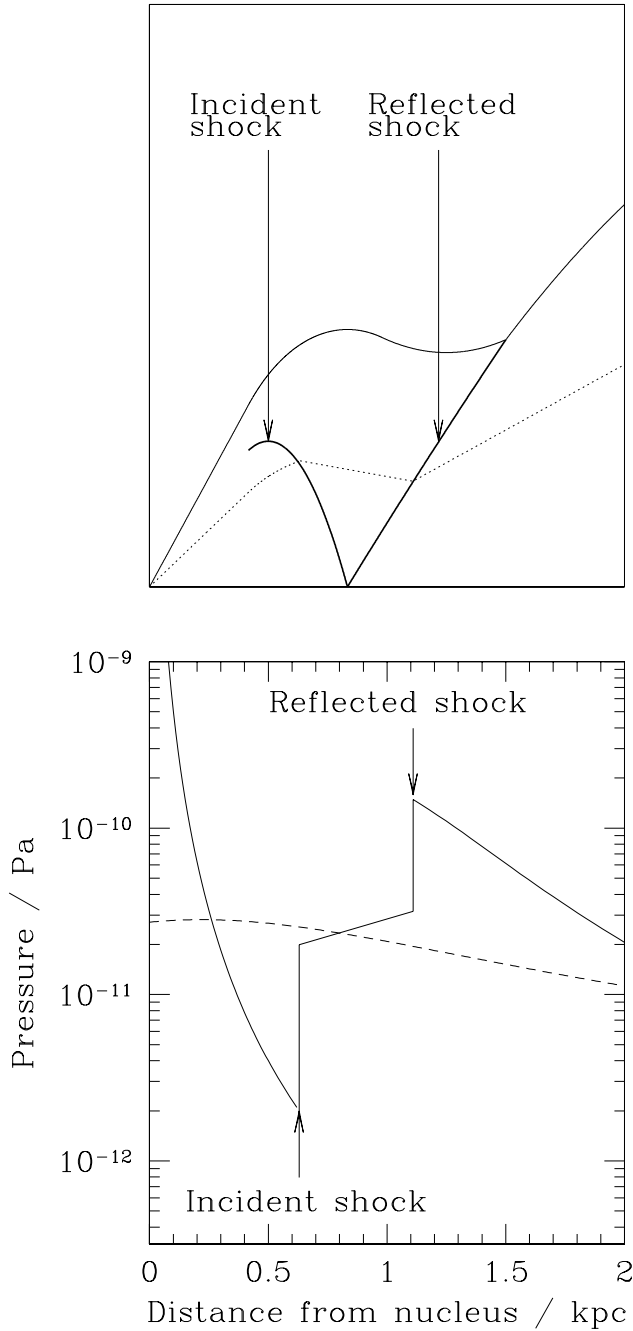}
\caption{The shock structure for a reconfining jet, after
\citet{Sand83}. The top panel shows a sketch with the edge of the jet
(full line), shocks (bold, full lines) and a representative streamline
(dotted line) marked.  Note that the vertical scale is expanded for
clarity. The bottom panel shows a sketch of internal (full) and external
(dotted) pressures against distance from the nucleus for the streamline in
the upper panel. The flaring point is set at 1.1\,kpc from the nucleus, as
for 3C\,31, the external pressure is as used in our models, as is the
internal pressure in the flaring region. All other quantities are notional.
\label{reconf}}
\end{figure}

We have also established that the flaring point in 3C\,31 is a discontinuity at
which the jet collimation, emissivity (and perhaps the velocity) change
abruptly and we have argued elsewhere that this is a general property of
FR\,I jets \citep{LPdRF}.  What causes this sudden transition? It has
frequently been suggested that it represents the onset of turbulence
\citep{Bic84} or (almost equivalently) the point at which Kelvin-Helmholtz
instabilities start to grow non-linearly \citep{RHCJ99,RH00}.  Our
requirement for a significant over-pressure at the flaring point leads us
instead to consider the possibility that the flaring point is associated
with a stationary shock system. The boundary position is
roughly consistent with the expected location of the reconfinement shock
formed when the internal pressure of a freely-expanding supersonic jet
falls below that of the external medium \citep{Sand83}.  Our estimates of
$p_{\rm sync}$ for the inner jet (Fig.~\ref{ext-fig}) are indeed
consistent with an over-pressure for the first 0.5\,kpc (but note that the
external pressure might be underestimated; \citealt{Hard_3C31}).  The shock is
expected to occur at a distance
\[ z_{\rm shock} \approx \left ( \frac{2 \Phi}{3 \pi p_{\rm ext} c} \right
)^{1/2} \approx \mbox{0.5 kpc} \] from the nucleus for a relativistic jet
\citep{Kom94}, in fortuitously good agreement with our results.
The  flaring point cannot be the initial
reconfinement shock (by definition, the jet recollimates at that point),
but the actual shock structure is likely to be more complicated. In the
calculations of \citet{Sand83}, a conical incident shock forms where the
jet has become significantly under-pressured. After this shock, the
internal pressure is still slightly below the ambient value.  The incident
shock is reflected off the jet axis to form a second conical shock, after
which the flow is over-pressured and expanding (Fig.~\ref{reconf}). It is
possible that the {\rm reflected} shock may represent the visible start of
the flaring region.  More detailed simulations will be required to
ascertain whether the over-pressure is consistent with the values inferred
earlier ($p/p_{\rm ext} \approx 8$), although \citet{Falle87} suggests
that values of $p / p_{\rm ext}$ as high as 12.5 are possible in a
non-relativistic jet if the oblique shocks are strong.

\subsection{Internal versus external entrainment}

Whilst the occurrence of a reconfinement shock provides a plausible
explanation for the over-pressure, it does not by itself explain the rapid
increase in mass input.  However: 
\begin{enumerate}
\item the increase in expansion rate will naturally lead to a larger mass
injection from stars, which will in turn expand the jet still further in a
runaway process and
\item the jet is expected to entrain the external medium more efficiently when it becomes
transonic.
\end{enumerate}
We estimate that internal entrainment from stars is within a factor of two of
that required to slow the jet over the first kiloparsec of the flaring
region (but note that the assumptions used to estimate the stellar mass
input are extremely crude, since they assume that the loss rates inside
and outside a jet are identical and the extent to which mass lost from
stars mixes with the jet is also poorly known).

A number of lines of evidence suggest, however, that external entrainment becomes
dominant further out. Firstly, our observations and kinematic model
\citep{LB02} show directly that there is an appreciable reduction in
velocity at the edges of the jet, as expected in external 
entrainment models.  The {\it shape} of the transverse
velocity profile in our best fit model changes relatively little down the
jet as it decelerates, so the profile could just be set close to the
nucleus.  An error analysis shows, however, that an evolution from a
top-hat velocity profile at the flaring point to a centrally-peaked
profile at larger distances would also be consistent with the data.
Secondly, the very sharp peak in the entrainment rate at a distance of
3.5\,kpc from the nucleus is difficult to reproduce with stellar mass
input alone: a maximum is indeed expected, but it should be much broader
(e.g.\ Fig~\ref{ent-ref}).  Thirdly, the appearance of locally low
polarization at the edges of the jets in the flaring region \citep{LB02}
requires the addition of radial magnetic field components at the edges of
this region, so that the field becomes almost isotropic there.
The appearance of such a field component suggests the onset of local
radial motions in this region, consistent with the inflow of ambient
material into the jet.  This effect occurs at exactly the position of the
local maximum of the entrainment rate (e.g.\ Fig.~\ref{ent-ref}).
Finally, the monotonic increase of entrainment rate at large distances is
clearly inconsistent with the fall-off in stellar density.  A general
feature of the models of \citet{Kom94} and \citet{BLK96}, where only
mass-loading from stars is considered, is that the jets are re-accelerated
at distances $>$1\,kpc, becoming significantly supersonic. The reason is
that the entrainment rate, which is proportional to the stellar density,
decreases rapidly.  Our results, which indicate a continuous deceleration,
therefore require additional entrainment, almost independent of the details
of the conservation-law analysis.

We conclude that external entrainment across the jet boundary from the galactic
atmosphere must be important in decelerating the jets,
but that internal entrainment from stars within the jet may play a significant role
in the initial phases.  Indeed, once the area of the jet starts
to increase, the mass input from stars may slow the flow to the point
where entrainment of external material becomes efficient \citep{Bic94}.

\subsection{External pressure and density}

The presence of a component of hot gas with a small core radius is
essential for the jets to decelerate without disruption. We have
confirmed, for example, that no solutions are possible with the
large-scale component associated with the NGC\,383 group alone. The core
radius ($r_c =$ 1.2\,kpc) and the distance of the flaring point from the
nucleus (1.1\,kpc) are almost exactly equal, so that the external pressure
gradient is steepest in the flaring region (Fig.~\ref{ext-fig}). We would
expect a significant external pressure gradient to drive the recollimation
of any flaring jet.  In one- or two-component beta-models of the type that
we have fit, this inevitably requires that the core radius (or one of the
core radii) be close to the flaring distance.

\subsection{Jet composition}

\subsubsection{Composition at the flaring point}

By comparing our estimate of jet density with the number of radiating
particles required to generate the observed synchrotron emissivity, we
can constrain the composition of the jet.  With the assumptions used
earlier to calculate the synchrotron minimum pressure, the number density
of radiating particles is:
\begin{equation}
n_{\rm rad} \approx 60 \gamma_{\rm min}^{-2\alpha}
\end{equation}
with $\alpha = 0.55$ if the power-law spectrum inferred from synchrotron
emission observed between 1.4 and 8.4\,GHz continues to lower energy.
This estimate uses the on-axis emissivity inferred for our best-fitting
kinematic model, which is close to the mean of the transverse profile.

The range of densities at the flaring point for pressure-matched models is
$\rho \approx$ 1.5 -- 3.5 $\times 10^{-27}$\,kg\,m$^{-3}$.  This
corresponds to $\gamma_{\rm min} \approx$ 50 -- 20 if every radiating
electron is associated with a proton.  This is a rough estimate whose
uncertainties include:
\begin{enumerate}
\item the derivation of $n_{\rm rad}$ from the emissivity, which assumes a
minimum-pressure condition;
\item the form of the spectrum at low energies, where we cannot observe
synchrotron radiation directly;
\item the range of densities derived from different models.
\end{enumerate}
If we drop the assumption of pressure balance in the outer region, the
constraints on $\gamma_{\rm min}$ are relaxed, but the only circumstance
in which we can avoid a low-energy cut-off entirely is if the
high-momentum-flux solutions are valid.

If, in contrast, the jet consists only of electrons and positrons at the
flaring point, then there would have to be a significant excess of
low-energy particles above the power-law extrapolation.

We conclude that, although the pressure-matched jets are very light, we
cannot exclude any of the following possibilities for their composition at
the flaring point:
\begin{enumerate}
\item relativistic electrons with a power-law spectrum with energy index
$2\alpha + 1 = 2.1$ and minimum Lorentz factor $\gamma_{\rm min} \approx$
20 -- 50, each accompanied by a proton;
\item an electron-positron plasma with some admixture of thermal matter,
the latter dominating the density;
\item a pure electron-positron plasma with an excess of particles over the
power-law prediction at low energies.
\end{enumerate}
Other intermediate compositions are possible.

\subsubsection{Entrainment in the inner jet}

Given that stellar mass input must occur in the inner jet, it is of
interest to estimate the mass flux at the flaring point due to this effect
alone.  This depends on knowledge of the luminosity density of the galaxy
at small radii, which is not available directly for NGC\,383 (HST optical
images show heavy dust obscuration, and infrared observations at
sufficiently high resolution are not yet available; \citealt{Vk}).  Given
the stellar luminosity of the galaxy, it is likely that the light profile
is of the ``core'' type, in which the surface-brightness profile shows a
break from a steep power-law at large radii ($\Sigma(r) \propto r^{-1.65}$
for NGC\,383; \citealt{Owe89}) to a shallow one ($\Sigma(r) \propto
r^{-a}$, with 0 \la $a$ \la 0.3) at small radii \citep{Lauer95}.  This
transition occurs around a break radius $r_b$ which is correlated with
absolute magnitude and is likely to be 100\,pc \la $r_b$ \la 1\,kpc for
NGC\,383 \citep{Faber97}.

Given the uncertainties, we have chosen to estimate two extreme limiting
cases for the mass input into the inner jet. In the first, we extrapolate
the $r^{-1.65}$ surface brightness profile seen at large radii inwards
from the flaring point. In the second, we assume a flat profile over the
whole of the inner jet, normalized at the flaring point (i.e.\ $r_b
\approx$ 1\,kpc and $a = 0$).  The number of particles injected per unit
volume per unit time is a Lorentz invariant, so we can derive the mass
flux at the flaring point by integrating the mass input rate
(equation~\ref{mdot-eq}) over the volume of the inner jet.  The results are:
$\Psi \approx 9.8 \times 10^{19}$\,kg\,s$^{-1}$ for $\Sigma(r) \propto
r^{-1.65}$ and $\Psi \approx 1.7 \times 10^{19}$\,kg\,s$^{-1}$ for a
constant surface brightness.

The predicted mass flux is at least commensurate with that estimated at
the flaring point (2.8 -- 3.4 $\times$ 10$^{19}$\,kg\,s$^{-1}$ for the
pressure-matched models). It is therefore possible that essentially {\em
all} of the mass of the jet comes from stars within
$\approx$1\,kpc of the nucleus. If the jet consists almost entirely of
electron-positron plasma on pc scales, it could still pick up enough mass
to be consistent with our estimates on kpc scales.  This argument is not
yet conclusive, because of the many uncertainties in estimating the
stellar mass input rate, but a jet consisting initially of pair plasma
would be entirely compatible with our results.

\subsubsection{Jet composition on parsec scales}

\begin{figure}
\epsfxsize=8.5cm
\epsffile{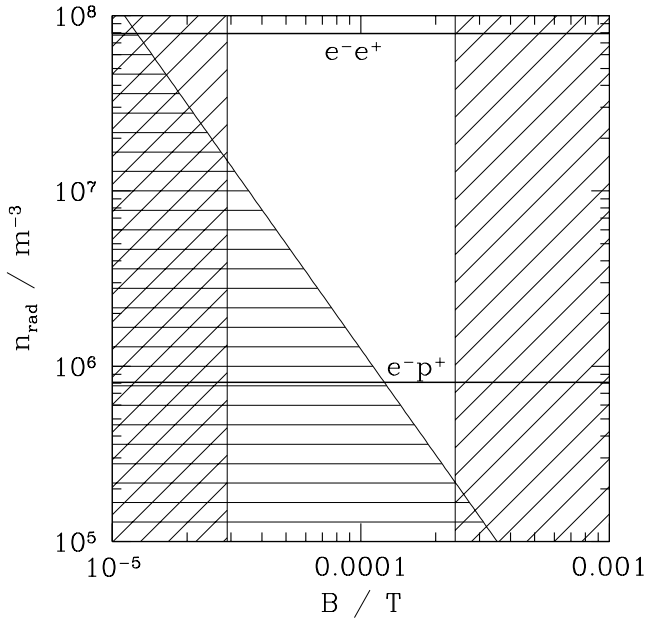}
\caption{Constraints on the $B$ -- $n_{\rm rad}$ plane imposed by
synchrotron self-absorption, total kinetic luminosity and synchrotron
self-Compton constraints, as in fig.\,1 of \citet{Rey96}.  Hatched areas
are excluded, as described in the text.  The two bold horizontal lines
represent the e$^-$e$^+$ and e$^-$p$^+$ cases for $\gamma_{\rm min} = 1$.
\label{Rey}
}
\end{figure}

We expect the amount of thermal material to increase from parsec scales to
the flaring point. We  therefore  compare our results with
those derived for pc scales using the methods of \citet{Rey96}.  These
authors used VLBI and X-ray observations of M\,87 to argue that its
parsec-scale jet is composed primarily of electron-positron plasma,
although they could not exclude an electron-proton jet with a low-energy
cut-off.  We have repeated their analysis for 3C\,31.  An upper limit to
the magnetic field strength is derived from the surface-brightness of the
self-absorbed core.  For an observing frequency of 4.973~GHz, an angular
diameter of $<$0.56~milliarcsec and a flux density of 0.071~Jy for the
core \citep{Lara97}, we deduce $B \la $ 2.4 $\times 10^{-4}$\,T for 3C\,31
if $\theta =$ 52.4$^\circ$.  Consideration of the absorption coefficient
at the point where the jet becomes optically thick gives $n_{\rm rad}B^2 >
0.02\gamma_{\rm min}^{-1} D_{\rm max}^{-2}$ where $n_{\rm rad}$ (in
m$^{-3}$) is the number density of radiating particles, $D_{\rm max}
= 1/\sin\theta = 1.26$ is the maximum Doppler factor and $B$ is in T.
Consequently, $n_{\rm rad}B^2 \ga 0.0125$ for $\gamma_{\rm min} = 1$.  We use the
value of the kinetic luminosity estimated earlier for the reference model
($\Phi = 1.1 \times 10^{37}$\,W) to solve for the particle number density
assuming e$^-$e$^+$ or e$^-$p$^+$ jets.  For a bulk Lorentz factor of 3,
as assumed by \citet{Rey96}, we derive $n_{\rm rad} \approx 7.5 \times
10^7$\,m$^{-3}$ for a pure e$^-$e$^+$ jet and $n_{\rm rad} \approx 8.1
\times 10^5$\,m$^{-3}$ for an e$^-$p$^+$ jet (note that $n_{\rm rad}$
consistently includes all radiating species).\footnote{In \citet{Rey96}'s
discussion of e$^-$p$^+$ jets, the quantity $n$ is used in different
places for the number densities of all radiating particles, and for
electrons alone.}  Finally, we deduce a lower limit to the magnetic field,
$B \ga 2.9 \times 10^{-5}$\,T using the X-ray core flux density at 1\,keV
from \citet{Hard_3C31} as an upper limit to the synchrotron self-Compton
emission.  The constraints are plotted in Fig.~\ref{Rey}.

The conclusions from this analysis are slightly weaker than those of
\citet{Rey96} for M\,87.  For 3C\,31, an e$^-$p$^+$ jet
with $\gamma_{\rm min} = 1$ just satisfies the constraints for a bulk
Lorentz factor $\Gamma = 3$, whereas it was formally ruled out for
M\,87. Our values for the jet composition and $\gamma_{\rm min}$ at the
flaring point would be consistent with the constraints shown in
Fig~\ref{Rey} even in the absence of any changes along the jet.

\section{Summary and further work}
\label{conclusions}

\subsection{Conclusions}

\subsubsection{Conservation-law solutions}

We have, for the first time, estimated the variations of pressure,
density, Mach number and entrainment rate along an extragalactic jet,
using a quasi-one-dimensional conservation-law approach combined with our
kinematic model and with measurements of the external gas properties from
{\em Chandra}.  Our solutions are physically self-consistent and satisfy
constraints set by the external and synchrotron minimum pressures. We
conclude that the deceleration law derived in \citet{LB02} is fully
consistent with the observed external density and pressure and the
conservation of energy, momentum and particles. The presence of a hot-gas
component with a small core radius, associated with the parent galaxy
nucleus rather than with the surrounding group, is required for the jet to
decelerate without disruption. 

All acceptable solutions have a number of common features:
\begin{enumerate}
\item The jets are overpressured by a factor of at least 4 the beginning of the
flaring region.  Thereafter, the pressure drops rapidly, approaching and
often falling below the external pressure.
\item At 3 -- 3.5\,kpc from the nucleus in the flaring
region, where the expansion rate is largest, there are local minima in the
pressure and density and maxima in the Mach number and entrainment rate.
\item Variations in the outer region are quite smooth: the density stays
roughly constant and the entrainment rate usually increases monotonically.
\item The Mach numbers are always $\approx$ 1 -- 2 in the outer region
(and usually also in the flaring region).
\end{enumerate}

\subsubsection{Outer region in pressure equilibrium; $\Pi = \Phi/c$}

It is very likely that the jets are close to pressure equilibrium
in the outer region and that the momentum flux $\Pi = \Phi/c$. The
principal uncertainty in our analysis is then the assumed velocity law,
which effectively integrates over the transverse distributions of velocity
for the relativistic and thermal plasma as functions of distance from the
nucleus.  We have explored a range of possible velocity laws, and conclude
that the energy flux must be in the range $9 \times 10^{36}$\,W \la $\Phi$
\la $1.4 \times 10^{37}$\,W.  At the flaring point, the key variables are:
\begin{description}
\item [{\bf density}] $\rho \approx$ 1.5 -- 3.5 $\times 10^{-27}$\,kg\,m$^{-3}$;
\item [{\bf pressure}] $p \approx$ 1.1 -- 1.8 $\times 10^{-10}$\,Pa;
\item [{\bf pressure ratio}] $p/p_{\rm ext} \approx$ 6 -- 9;
\item [{\bf Mach number}] ${\cal M} \approx$ 1 -- 2;
\item [{\bf mass flux}] $\Psi \approx$ 2.8 -- 3.4 $\times 10^{19}$\,kg\,s$^{-1}$;
\item [{\bf entrainment rate}] $\approx$ 1.0 -- 1.3 $\times 10^{19}$\,kg\,kpc$^{-1}$\,s$^{-1}$;
\item [{\bf kinetic/internal energy ratio}] ${\cal R} \approx$ 0.30 -- 0.44.
\end{description}
The variations of these quantities along the jets are illustrated in
Fig.~\ref{consfit}.  Uncertainties in the kinematic model
(Section~\ref{vel-errors}) and the external pressure and density
(Section~\ref{ext-var}) lead to similar, but smaller, variations in the derived
flow variables.

\subsubsection{Outer region not in pressure equilibrium; $\Pi = \Phi/c$}

If the jets are not in pressure equilibrium in the outer region, then a
wider range of energy fluxes is allowed.  The lower limit ($4.5 \times
10^{36}$\,W) is set by the condition that the internal pressure exceeds
the synchrotron minimum; the upper limit either by the requirement that
the mass flux always increases away from the nucleus or by an
over-pressure constraint ($3.0 \times 10^{37}$\,W or $2.3 \times
10^{37}$\,W, respectively).  There is a wider range of solutions
(Fig.~\ref{consfita}), but they are qualitatively similar to those for the
pressure-matched case.

\subsubsection{Synchrotron minimum pressure and the equation of state}

Our conclusions are not seriously affected by changes in the assumptions used to
derive the minimum pressure from the synchrotron emissivity
(Section~\ref{sync-press-var}).  Our assumption that the internal energy of the
jets is dominated by relativistic particles and magnetic field is
self-consistent, but we have also shown that our results are insensitive to
changes in the equation of state (Section~\ref{eos}).

\subsubsection{$\Pi \neq \Phi/c$}
\label{mom-flux-summary}

If the momentum flux is allowed to exceed $\Phi/c$ by a large factor, the
solutions are much less well constrained (Section~\ref{mom-flux-var}). The jets
can be much denser, and entrain more rapidly than those having $\Pi = \Phi/c$
(Fig.~\ref{consfitb}).  Although we cannot rule out these solutions from our
data alone, they are incompatible with the need to decelerate from high Lorentz
factors on parsec scales and require extremely high entrainment rates even where
the jet Mach number ${\cal M} \approx 5$.  We suggest that these solutions are
highly unlikely.

\subsubsection{The deceleration mechanism}

The large over-pressure at the beginning of the flaring region suggests
the presence of a stationary shock, perhaps associated with reconfinement
of the jet.  The amount of mass lost by stars inside the jets and the
degree of mixing of the ejecta are both very uncertain, but our best
estimate is that stellar mass input is within a factor of two of the rate
needed to slow the jet at the beginning of the flaring region. At larger
distances, the required entrainment rate is much higher than could be
supplied by stars and also increases with distance from the nucleus in a
region where the stellar density falls rapidly.  We conclude that another
mass source (presumably entrainment from the large-scale galactic 
atmosphere across the boundary layer of the jet) must dominate at large
distances; perhaps everywhere, but that stellar mass input could still
significantly affect the initial deceleration. A second piece of evidence
in favour of entrainment of external gas across the jet
boundary is the (approximate) isotropy of the magnetic
field at the edge of the flaring region \citep{LB02}, which is most easily
interpreted as the effect of disordered motions in a turbulent 
entraining flow.

\subsubsection{Jet composition}

Our estimate of stellar mass injection within 1\,kpc of the nucleus is most
consistent with the hypothesis that the jets consist primarily of pair plasma on
parsec scales and that most of their mass at the flaring point is in the form of
entrained thermal plasma.  A jet consisting entirely of electron-positron plasma
at the flaring point would require a very large low-energy excess over a
power-law energy spectrum. Given the uncertainties in our estimates, we cannot
rule out an electron-proton composition; this would require a minimum Lorentz
factor of $\gamma_{\rm min} \approx$ 20 -- 50 for the radiating electrons.

\subsection{Further work}

\subsubsection{Observations}

The next step in this work is to carry out kinematic modelling and X-ray
observations of other sources and to investigate how the jet behaviour depends
on galaxy properties and luminosity.  Particularly important questions include:
\begin{enumerate}
\item Is flaring and recollimation always associated with a steep external
pressure gradient?
\item Is there a difference in the entrainment rate for sources whose jets
propagate entirely within their radio lobes (presumably much less dense
than the external medium) compared with those, like 3C\,31, where the jets
appear to be in direct contact with the hot gas?
\item How does the deceleration process depend on jet power?
\item What is the stellar density close to the nucleus? (This will require
infra-red imaging at high spatial resolution).
\item Is there morphological evidence for the reconfinement shock
structure we have suggested?
\item What limits can we set on the energy spectrum of the
relativistic electrons from low-frequency radio and high-frequency
(optical -- X-ray) observations?
\item Can we refine the constraints on pc-scale jet composition by
higher-resolution VLBI imaging or measurements of circular polarization
\citep{Ward}?
\end{enumerate}

\subsubsection{Theory}

Our analysis also poses a number of challenging theoretical problems:
\begin{enumerate}
\item Is it possible to simulate entrainment into a decelerating,
relativistic, magnetized jet with the very low density contrast we infer
and in a realistic galactic atmosphere?
\item Is the required over-pressure at the flaring point consistent with
the shock structure in a reconfining jet?
\item What is the viscosity mechanism? How is momentum transported across
the jet?  How can we constrain this using estimates of the velocity profile?
\item Are turbulent velocities significant? What are their effects on
energy and momentum transport and magnetic fields? 
\item Is an ultra-relativistic equation of state an adequate approximation
everywhere?
\item How is the entrained material mixed and heated?
\item Can better estimates be made of the mass input rate from stars
inside a jet?
\end{enumerate}

\section*{Acknowledgments}

RAL would like to thank the National Radio Astronomy Observatory, the
Istituto di Radioastronomia, Bologna and Alan and Mary Bridle for
hospitality during this project.  We acknowledge travel support from NATO
Grant CRG931498.  The National Radio Astronomy Observatory is a
facility of the National Science Foundation operated under cooperative 
agreement by Associated Universities, Inc.

\end{document}